\documentclass[prd,showpacs,preprintnumbers,floatfix]{revtex4}

\usepackage{axodraw}

\newcommand{\be}{\begin{equation}}
\newcommand{\ee}{\end{equation}}
\newcommand{\bea}{\begin{eqnarray}}
\newcommand{\eea}{\end{eqnarray}}
\newcommand{\nc}{N_c}

\begin{document}

\preprint{UCD2002-3/RU-02-01}
\title{Proposal for the numerical solution of planar QCD}
\author{J. Kiskis}
\affiliation{Department of Physics, University of California,
Davis, CA 95616, USA}
\author{R. Narayanan}
\affiliation{ American Physical Society,
One Research Road, Ridge, NY 11961, USA}
\author{H. Neuberger}
\altaffiliation[Permanent Address: ]{Rutgers University, Department of Physics 
and Astronomy, Piscataway, NJ 08855, USA}
\affiliation{
Department of Physics, Technion - Israel Institute of Technology\\
Haifa 32000, Israel\\
School of Physics and Astronomy\\
The Raymond and Beverly Sackler Faculty of Exact Sciences\\
Tel Aviv University, 69978 Tel Aviv, Israel\\
Laboratoire de Physique Th\'eorique de l'Ecole 
Normale Sup\'erieure\\
24 rue Lhomond, 75231 Paris Cedex, France}

\begin{abstract}
          
Using quenched reduction, we propose a method for the numerical calculation of
meson correlation functions in the planar limit of QCD. General features of the
approach are outlined, and an example is given in the context of 
two-dimensional QCD.

\end{abstract}


\maketitle

\section{Introduction}

The phenomenological appeal of planar QCD is well known \cite{coleman}. It is 
believed that a 
limit in which the number of colors $N_c$ goes to infinity while the coupling
goes to zero with fixed $g^2N_c$ produces the simpler planar theory while still
preserving essential perturbative and non-perturbative features of the physical
$N_c=3$ case \cite{thooft,thooft2d}. However, an analytical description of the
planar limit has remained elusive. This has left lattice gauge theory as the
prevailing approach for the conceptual and quantitative analysis of strong
coupling in QCD.

Recent numerical investigations with increasing $N_c$ \cite{teper,federico}
indicate that some $N_c=3$ physical quantities are not far from their 
$N_c=\infty$ limits. This suggests that numerical results from planar QCD might
have phenomenological value. However, in numerical lattice calculations, it is
easier to deal with the $N_c=3$ case than to approach the planar limit by
simply increasing $N_c$. An approach that arrives at the simpler planar theory
more directly is indicated. The Eguchi-Kawai \cite{ek} reduction offers such a
shortcut. It has the potential to require much less computation because the
planar limit of some observables is obtained from a matrix model with only the
$d \times N_c^2$ degrees of freedom of $d$ unitary matrices. Since the
dimension $d$ enters linearly rather than exponentially, it seems that four
dimensions has only twice the computational cost of two dimensions.

Using the Eguchi-Kawai reduction to approach the planar limit of QCD by first
going to infinite $\nc$ and then to zero lattice spacing might be beneficial
because the two limits may not commute \cite{neub,moshe}. Also the
commutativity of the infinite volume and infinite $\nc$ limits might be
questioned because gauge theories are massless, and the construction of
\cite{thooft2} does not apply. We begin with the assumption that meson-meson
correlation functions at non-zero momenta are ``infrared  safe''. If 
our calculations in two dimensions are finite and agree with analytical
results, then we can be confident about calculations in four dimensions,
where infrared problems are less serious.

In the Eguchi-Kawai model the entire $d$-dimensional 
lattice is reduced to a point. Nevertheless
the matrix model incorporates the lattice structure. This is best understood 
in momentum
space \cite{bhn}, where the $d$ sets of eigenvalues of the link matrices 
populate  
the momentum $d$-torus. To ensure that the torus is indeed 
covered, one needs to alter the Eguchi-Kawai
proposal. The first suggestion for how to achieve this was to freeze 
the eigenvalue variables and remove them from the integration over 
gauge fields. The fluctuations of the remaining gauge degrees of
freedom are not allowed to back-react on the eigenvalue distribution.
The expectation value of gauge invariant observables
is subsequently averaged over different sets of angles. This produced the
quenched Eguchi-Kawai model \cite{bhn}. 
Related work on quenched reduced models appeared 
in~\cite{grosskita,migdal,parisi}.
A more elegant method, which involves a change in the action by a 
``twist'' \cite{tek}, was subsequently discovered.
This clever approach has also 
led to the discovery of non-commutative gauge field theories \cite{noncomm} 
starting from the lattice.

Nevertheless our proposal is in the quenched version \cite{bhn}
because the introduction of fermions in the
twisted case is somewhat restricted. 
The addition of fermions to reduced models was first
carried out in a continuum version in \cite{grosskita}. A somewhat 
similar procedure on the
lattice was introduced in \cite{neublevine1,kazmig1}. 
In the latter version, the number of flavors $N_f$
was a free parameter and 
could be taken to infinity. This produces a reduced model leading 
to the Veneziano limit \cite{venez} rather than to 
the 't Hooft limit \cite{coleman}. Following
up on the idea of \cite{neublevine1}, 
a twisted version of the $N_f=N_c$ case was presented in \cite{twivenz}. Since we
want to have a variable number of flavors, we opt for the quenched version. In
addition, the quenched version is computationally simpler than
the twisted one even for $N_f = N_c$ because the Dirac operator is block-diagonal
with $N_f$ blocks each of size $N_c\times N_c$. In the twisted case, one has a full
$N_c N_f\times N_c N_f$ matrix. 

Our proposal specifically targets mesonic observables.
To get glueball masses from a reduced model, we would need to keep one direction
unreduced as described in \cite{neublevine2}. 
To investigate finite temperature, we would again
need a Hamiltonian-like formulation \cite{neubham}. 
In addition to the calculation of meson masses,
our approach allows for the study of fermion quenching effects
(valence approximation effects) and finite fermion number 
density at zero temperature. 
Also the pathologies of the valence approximation should disappear in the 
limit of large $N_c$ with fixed $N_f$.
The fermionic bilinears which define
the meson fields are the building blocks of the four-fermi operators crucial
to the study of weak decays. We envisage taking the $\frac{1}{\nc}$ approach
to non-leptonic weak decays \cite{bardeen} to the lattice where it could be
compared to more direct approaches and where some of its phenomenological input
could be replaced by numerical data obtained from ``first principles''. 

There have been both practical and theoretical developments that motivate a
return to numerical work on planar QCD.
Theoretical progress has led to the new lattice fermions \cite{lattchiral},
which might be able to solve some problems encountered about twenty years ago 
\cite{grosskita}.
The problems had to do with the meaning of topological charge density in the
reduced model, chiral anomalies, and the $\theta$-parameter dependence of the 
vacuum action density in the planar limit \cite{witten}. 
Equally important are developments in computer hardware.
Since the initial numerical experiments on reduced models for planar QCD,
computers have become $10^4$ times more powerful
for fixed power consumption.
We suggest that reduced model simulations are a worthy job for 
weakly (and cheaply) linked PC clusters. 
Moreover at infinite $\nc$, the Wilson Dirac operator in the
argument
of the sign function has no small eigenvalues when the pure gauge action of
the reduced
model is of the usual single plaquette type, and the rescaled lattice gauge
coupling is large
enough (but not infinite). 
This is unlike the situation in ordinary QCD~\cite{smallev}.
Thus the numerical implementation of the overlap
Dirac
operator will be simpler than in ordinary QCD.

In the next section,
we define reduced models starting from ordinary formulations of lattice gauge theories.
We show that the 
topological charge remains a nontrivial object even after
reduction and that reduced models have, in some sense, more topological numbers
than their unreduced counterparts. We briefly discuss chiral gauge theories. The
meson-meson correlators, which are the most practical aspect of our
proposal, are also introduced. Section three is focused on the
special case of two dimensions, in which we test the numerical methods. 
Although
planar QCD$_2$ has been solved exactly in the weak coupling phase,
additional analytical and numerical work is needed to extract the results that
can be compared to Monte Carlo data. Although there is a vast literature on
QCD$_2$, in the interest of clarity, 
we have included a rather detailed account in
the Appendix. In the summary section, we describe
some of our ideas for future developments. We believe
that there are many options for future research and that the
potential for truly new and interesting results is quite high. 

\section{General Framework}
\subsection{Original lattice model and its topological charge}
\vskip .3in
Before reduction, the lattice systems of interest are defined on a 
$d$-dimensional hypercubic lattice with a total $V$ sites on a $d$-torus of 
equal sides. Sites are labeled by $x$ and positive directions by 
$\mu = 1,2,\dots, d$. There are $SU(\nc)$ gauge fields
$U_\mu (x)$ on the links, and fermions interact with this
gauge background. 

Since we are going to deal later with fermions in the fundamental
representation, it is best to introduce the fields now, even though
we are first focusing on the pure gauge sector. 
The fermion field $\psi$ is a vector in $V_s \otimes V_c \otimes V_l$, the
tensor product of the spin, color, and lattice vector spaces, respectively.
Its components are $\psi_\alpha^i (x)$ with spin index $\alpha$,
color index $i$ and lattice position $x$.
There are several operators acting on fermions:
The Euclidean Dirac $\gamma_\mu$'s
act only on spinorial indices, $\gamma_\mu : V_s \to V_s$. 
The directional parallel transporters $T_\mu$
act on the site index and the group index ($T_\mu : V_c\otimes V_l \to
V_c \otimes V_l $) as follows:
\be
\label{tmu}
T_\mu (\psi ) (x) = U_\mu (x) \psi (x+\hat\mu) .
\ee
For $U_\mu (x) = 1$, the $T_\mu$ become
commuting shift operators $T_\mu^0$. For $V=L^d$ each $T_\mu^0$ is
unitary and has eigenvalues $e^{\frac{2\pi\imath}{L} k},~k=0,1,2,\dots,L-1$.
Hence for even $d$, $T_\mu$ is not only unitary but also has unit determinant
($\det T_\mu^0 = (-1)^{d(L-1)}$).  

Gauge transformations
are characterized by a collection of $g(x)\in SU(\nc )$
acting on $\psi$ pointwise, and only on the group
indices. The action is represented by a unitary operator 
$G(g)$ : $V_c\otimes V_l \to
V_c\otimes V_l$ with $(G(g)\psi)(x) =g(x)\psi(x)$. The $T_\mu$ operators 
are gauge covariant
\be
\label{ginv}
G(g) T_\mu (U) G^\dagger (g) = T_\mu (U^g ) 
\ee
with
\be
U^g_\mu (x) = g(x) U_\mu (x) g^\dagger (x+\hat\mu ) .
\ee
The variables $U_\mu (x)$ are distributed according to a
probability density that is invariant under $U\to U^g$ for any $g$.

For the commutators $[T_\mu , T_\nu ]$, we have
\be
[T_\mu , T_\nu ]^\dagger [T_\mu , T_\nu ]= (1-P_{\mu\nu})^\dagger
(1-P_{\mu\nu})=2-P_{\mu\nu}-P_{\mu\nu}^\dagger
\ee
with the unitary $P_{\mu\nu}$ given by
\be
P_{\mu\nu} = T_\nu^\dagger T_\mu^\dagger T_\nu T_\mu   .
\ee
The operators $P_{\mu\nu}$ are site diagonal
with entries that are the parallel transporters around plaquettes:
\bea
&(P_{\mu\nu}\psi)(x) = U_{\mu\nu} (x) \psi (x)\\
&U_{\mu\nu}(x)=
U_\nu^\dagger (x-\hat\nu ) U_\mu^\dagger (x -\hat\nu -\hat\mu ) 
U_\nu (x -\hat\nu - \hat\mu ) U_\mu (x-\hat\mu ) .
\eea
$U_{\mu\nu} (x)$ is associated with the elementary loop
starting at site $x$, going first in the negative $\nu$ direction,
then in the negative $\mu$ direction, and coming back round the plaquette. 

The Wilson pure gauge action is given by
\be
S_g = \beta \sum_{\mu > \nu} Tr (P_{\mu\nu} + P_{\mu\nu}^\dagger )=
{\rm Const} - \beta \sum_{\mu > \nu} Tr [T_\mu,T_\nu ]^\dagger [T_\mu,T_\nu ].
\ee
In the continuum limit, configurations with all the $U_{\mu\nu}(x)$ 
close to the unit matrix are preferred.

All our operators
are finite dimensional matrices. 
The norm of $A$ is $\| A\|$ with
$\| A\|^2$ the largest
eigenvalue of $A^\dagger A$.
We have
\be
\|[T_\mu , T_\nu ]\| = \|1 - P_{\mu\nu}\|  .
\ee

The gauge invariant constraint  
\be
\label{rest}
\|[ T_\mu , T_\nu ]\| \leq \epsilon_{\mu \nu} 
\ee
for all $\mu > \nu$ does not change the continuum limit.
It is equivalent to
\be
\|1-U_{\mu\nu}(x)\| \leq \epsilon_{\mu\nu} 
\ee
for every site $x$. The $\epsilon_{\mu\nu}$ are small fixed nonzero numbers,
independent of $\beta$ and of $\nc$. The probability distribution of the
link variables is $e^S$ with integration ranges restricted by (\ref{rest}).

The lattice version of the massive continuum Dirac operator $D(m)$:
$V_s\otimes V_c\otimes V_l\to V_s\otimes V_c \otimes V_l $
is an element in the algebra generated 
by $T_\mu,~ T_\mu^\dagger ,~ \gamma_\mu$ and thus is 
gauge covariant. 
The Wilson Dirac operator
$D_W (m)$ is the sparsest possible analogue of the continuum massive
Dirac operator which obeys hypercubic symmetry. 
Fixing the $r$-parameter
to its preferred value ($r=1$),
\be
\label{dw}
D_W = m+\sum_\mu(1- X_\mu )~~~~~X_\mu^\dagger X_\mu =1~~~~X_\mu=
{{1-\gamma_\mu}\over 2} T_\mu +{{1+\gamma_\mu}\over 2} T_\mu^\dagger .
\ee

One can then prove a general bound ~\cite{bound}:
\be
\label{bnd}
\left [ \lambda_{\rm min} (D_W^\dagger (m ) D_W (m))\right ]^{1\over 2}\geq
\left [ 1 - (2+\sqrt{2}) 
\sum_{\mu > \nu }\epsilon_{\mu\nu} \right ]^{1\over 2} -|1+m| .
\ee
Here $\lambda_{\rm min}({\rm M})$ is the smallest eigenvalue of the matrix
${\rm M}$. 
This bound has content only for $m$ close to $-1$:
\be
|1+m|\leq \left [ 1 - (2+\sqrt{2}) 
\sum_{\mu > \nu }\epsilon_{\mu\nu} \right ]^{1\over 2}  .
\ee
This range is contained in the open interval
$-2 <  m < 0 $.

For even $d$, we define the hermitian operator 
\be
H_W (m) = \gamma_{d+1} D_W (m)~~~~\text{for which}~~~~~H_W^2 (m) = 
D_W^\dagger (m) D_W(m) .
\ee
It follows from (\ref{rest}) and (\ref{bnd}) that $H_W$ has no zero 
eigenvalues so that its sign function $\epsilon \equiv {\rm sign}(H_W) $ is 
well-defined~\cite{overlap}.
The topological charge of a gauge field configuration is given by
\be
\label{top}
Q=\frac{1}{2} Tr \epsilon
\ee
The value of $Q$ cannot be changed by a smooth deformation that respects the
bounds (\ref{rest}). 
Close to continuum, $Q$ equals the continuum
topological charge \cite{adams}. The trace in (\ref{top})
includes a sum over lattice sites. Leaving this
sum out produces a local functional of the lattice gauge background
which is a lattice version of the topological charge density $trF\tilde F (x)$.
For a review, see \cite{dubna}. 

It is important that the proof~\cite{bound} 
of the bound makes no use of the special
structure of the matrices $T_\mu$ in (\ref{tmu}) 
except their unitarity. Thus
any unitary $T_\mu$ with
commutators close to zero (\ref{rest}) can be used in (\ref{dw}) and 
(\ref{top}) to define homotopy classes for the gauge backgrounds.
Of course, the ``local'' topological density
has no meaning if the $T_\mu$'s are that general.

Although we spoke about fermions and used gamma matrices, they were
auxiliary concepts. We only dealt with the pure gauge sector. The gamma
matrices played a role in defining the topological number, but
the latter is just a function of the gauge field background. 
It will be important in the following discussion that this definition of $Q$ 
can be non-zero on a finite lattice.
One need not to go to an infinite lattice to see non-zero topology. 

\subsection{Reduced model and reduced topological charge}

The reduced model for lattice gauge fields~\cite{ek} is defined on
a toroidal lattice with a single site. The matrices
$T_\mu$ become unitary matrices with no restrictions on their structure
since they are equal to $U_\mu$ (cf. (\ref{tmu})) on a single site lattice.
A connection between models differing only in the number of sites $V$
can be established by taking $\nc$ to infinity with $\beta/\nc$
fixed. The leading behavior of traces of Wilson loops written in terms of 
the $T_\mu$
is then independent of $V$. Actually this is only true for $\beta/\nc$
small enough. It fails for large $\beta/\nc$ and thus also for the continuum
limit ~\cite{bhn}. The key to fixing the model is to
understand that the $V$ independence is caused by momentum space getting
absorbed into the space of eigenvalues of the operators $T_\mu$. Consequently,
the quenched reduced lattice model is defined by 
freezing the eigenvalues of $T_\mu$ to 
a set uniformly distributed over the unit circle for each $\mu$
\cite{bhn}.  
\begin{equation}
T_\mu = V_\mu D_\mu V_\mu^\dagger,~~~~~~D_\mu={\rm diag} (e^{i\theta_\mu^1},
e^{i\theta_\mu^2},\cdots,e^{i\theta_\mu^{\nc}})
\label{tmured}
\end{equation}
The angles $\theta_\mu^i$ are randomly distributed in the interval
$[-\pi, \pi ]$. The ordering of the angles for each direction 
$\mu$ does not matter since
one can go from one ordering to another by appropriately changing the
integration variables $V_\mu$. In other words, only the sets of eigenvalues are
frozen, not the ordering within each set. 
The matrices $V_\mu$ are unitary and otherwise restricted
only by (\ref{rest}). 

The pure gauge action is invariant under $V_\mu \to V_\mu D^\prime_\mu$ where
the $D_\mu^\prime$ are arbitrary diagonal unitary matrices.
Thus the $V_{\mu}$ should be thought of as
taking values in the coset space $U(\nc)/(U(1))^{\nc}$. 
The most important symmetry of the pure gauge action is
gauge invariance which now acts by conjugation  
$T_\mu \to \Omega T_\mu \Omega^\dagger$. 
One can think of $\Omega$ as the replacement for $G(g)$ in (\ref{ginv}). The
gauge group is $SU(\nc)/Z(\nc)$. 

The absolute minima of the action are given by $T_\mu$'s that are 
simultaneously diagonal. These minima are
changed one into another by certain sets of $V_\mu$ matrices. In the double
line notation \cite{bhn,grosskita,migdal,parisi} of planar graphs, the crystal 
$d$-momentum going through a line is given, component by component,
by the difference between the angular 
phases of the two eigenvalues of $T_\mu$ associated with 
each of the color indices attached to each component of the double line. 
Different minima give different momenta, and the sum over
minima together with the sum over color indices reproduces the ordinary
planar lattice Feynman diagrams. As the
typical spacing between a $\theta_\mu^i$ and the closest 
$\theta_\mu^j$ is of order $1/\nc$,
it follows that the momenta values are in effective correspondence with those 
in a finite volume of size $V={\nc}^d$. 
On the other hand, focusing on a single minimum might suggest
$V=\nc$. In the twisted case, one typically gets $V={\nc}^{\frac{d}{2}}$. 
In numerical simulations, 
it is important to ensure that all minima are properly sampled.

In the continuum version of the reduced model~\cite{grosskita},
the basic variables are traceless hermitian matrices $A_\mu$ with
restrictions on their eigenvalues. One could try a ``geometric''
definition of topological charge, but it does not work well~\cite{grosskita}.
The natural expression corresponding to the non-abelian field strength is 
$F_{\mu\nu} = [A_\mu , A_\nu ]$. So long as the $A_\mu$ are finite matrices,
\be
Tr F\tilde F \propto \epsilon_{\mu\nu\rho\sigma} Tr A_\mu A_\nu A_\rho A_\sigma =0
\ee
because of the cyclicity of the trace and the antisymmetry of the epsilon symbol.
To get non-zero topological charge, one needs to work with infinite 
$A_\mu$ matrices, and a numerical approach seems impossible. 
The topological
charge of the lattice reduced model 
is defined by (\ref{top})
using unitary $T_\mu$'s that are restricted only by (\ref{rest}).
As already mentioned, the definition of topological
charge and the restriction on the commutators (\ref{rest}) 
work exactly the same
way as before. 

\begin{figure}
\epsfxsize = 0.8\textwidth
\centerline{\epsfbox{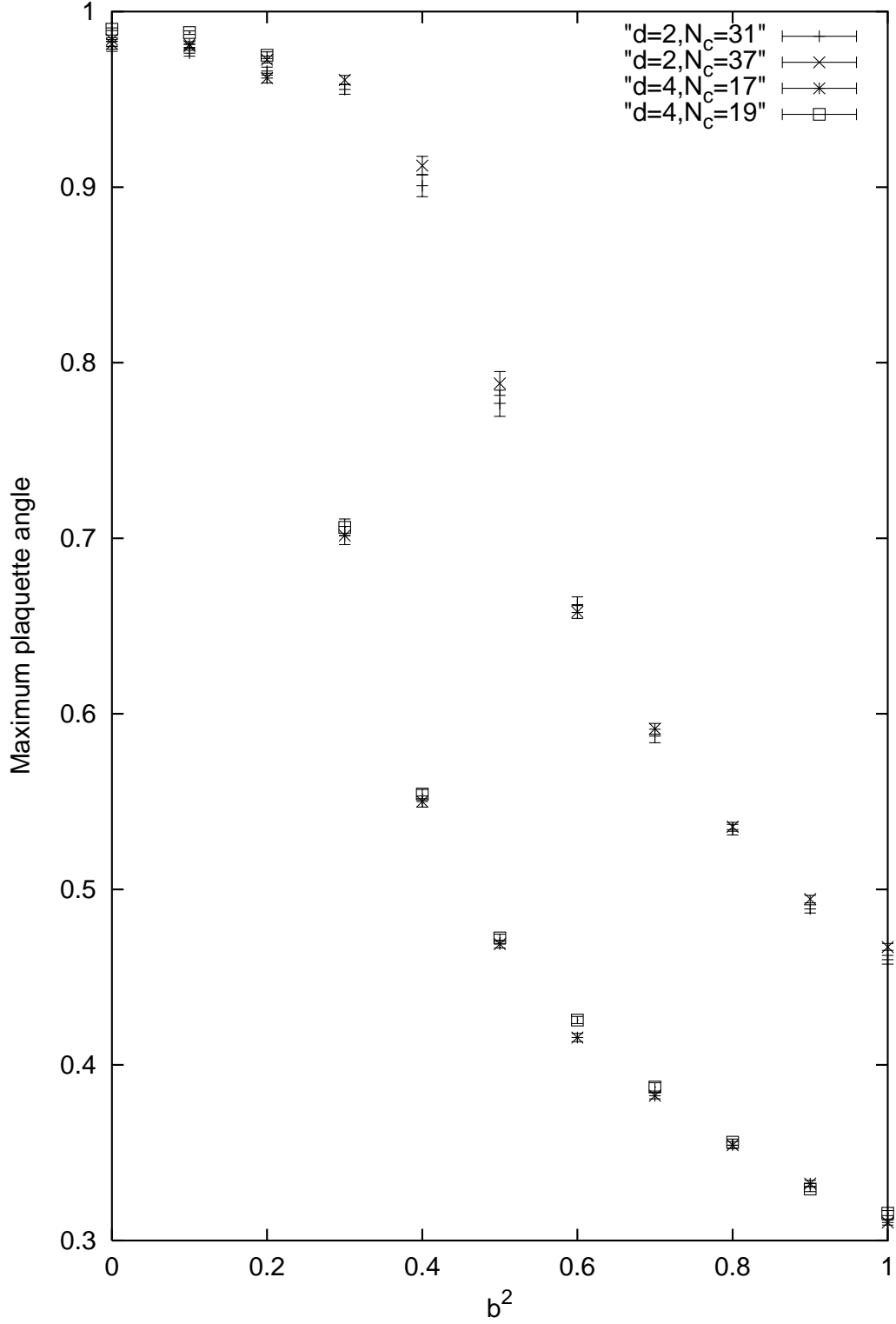}}
\caption{
Plot of the maximum plaquette angle in units of $\pi$   
as a function of
$b^2={\beta\over N_c}$. 
The data is at $N_c=31$ and $N_c=37$ for the two dimensional model
and is at $N_c=17$ and $N_c=19$ for the four dimensional model.
}
\label{plaq}
\end{figure}

This extra constraint (\ref{rest}) on the commutators of $T_\mu$ 
is not necessary at
infinite $N_c$. For $d=2$, when we get close enough to the continuum limit
and the lattice gauge coupling $b^2=\beta/N_c$ is large enough, we
are far below the Gross Witten phase transition~\cite{gw},
and the plaquette eigenvalue distribution
is contained in a small neighborhood of unity.
This is evident in Fig.~\ref{plaq} where 
we have plotted the expectation value of the maximum angle of the untraced 
plaquette in
units of $\pi$ for two different values of $N_c$ in the quenched model. 
This is also true 
in four dimensions where again there is
a phase transition as seen in the Fig.~\ref{plaq}.
A similar result was also found in the four dimensional twisted
Eguchi-Kawai model~\cite{twistedplaq}.
The numerical evidence strongly indicates that
the extra condition
restricting the matrices is satisfied as long as we are close enough to 
continuum even if we use only the simplest
single plaquette Wilson action.

A complete study of topological charge is beyond the scope of this paper
and our discussion will be brief.
Consider a two dimensional
gauge field background in which the $T_\mu$'s 
represent a $U(1)$ background that is a $U(1)$ instanton on a
two dimensional torus. The $T_\mu$ have the structure of a ``shift'' and a
``clock'' matrix \cite{recent}
\begin{equation}
\label{Xmatrix}
T_1\equiv X =\pmatrix{0 & 1 & 0 & \dots & 0 \cr
                  0 & 0 & 1 & \dots & 0 \cr
                  \dots  & \dots &\dots &\dots & \dots   \cr
                  0 & 0 & \dots & 0 & 1 \cr
                  1 & 0 & \dots & 0 & 0  }
\end{equation}
\begin{equation}
\label{Ymatrix}
T_2\equiv Y = \pmatrix{1 & 0 & 0 & \dots & 0 \cr
                   0 & \zeta & 0 & \dots & 0 \cr
                   0 & 0 &\zeta^2 & \dots & 0 \cr
                  \dots  & \dots &\dots &\dots & \dots   \cr
                   0 & 0 & \dots & 0 & \zeta^{N-1}  }   .
\end{equation}
Here $\zeta=e^{\frac{2\pi\imath}{\nc}}$, and we take $N=\nc$.  
This abelian background on a lattice of area $\nc$ can be viewed as
a non-abelian $SU(\nc )$ background in the reduced model. As far as
the spectra of the various lattice Dirac operators go, how we look at
the background is irrelevant. Hence, the reduced model has 
$Q=1$ for this configuration.
Moreover calculating the gauge field action, we easily can check that
it stays finite in the 't Hooft large $\nc$ limit. Similarly,
one can easily check that (\ref{rest}) is satisfied for $\nc$ beyond some
finite value. 

One expects no
topological effects in two dimensional planar QCD. This is not
in contradiction with our observation 
because the continuum limit is obtained when the fixed
parameter of the 't Hooft limit $\beta/\nc$ is taken to infinity,
and these backgrounds do not contribute. 
We learn that lattice reduction may perform a ``miracle'' in the sense
that one may find topological effects that were not present in the original
continuum model.  Whether or not these effects are relevant in the continuum
limit is a dynamical question. 

What happens in four dimensions? We have seen that what looks like an abelian
background before reduction can appear as a nonabelian background in the 
reduced model. (Actually, the results of \cite{review} say that for any finite
$\nc \ge 3$
on a continuous torus, 
any allowed value of $Q$ can be attained by an abelian background. We are not
making use of that result here.) 
In four dimensions, one can obtain a configuration that
is nontrivial topologically, for example, by choosing
an abelian instanton in the $1-2$ plane and another abelian instanton in the $3-4$
plane. These instantons would lead to shift-clock pairs as above. We could take
$\nc=N^2$ and write the $T_\mu$'s as direct products of $N\times N$ matrices:
$T_1=X\otimes 1$, $T_2=Y\otimes 1$, $T_3 = 1 \otimes X$ and $T_4=1 \otimes Y$. 
All these matrices are in $SU(N^2 )$. 
The topological charge of this background is $Q=1$. The spectra of the $T_\mu$'s
are highly degenerate in this example. To produce truly acceptable configurations
in a quenched reduced model, we would need to find deformations of these matrices
that lift the degeneracy while preserving $Q=1$. 

The action of this configuration diverges linearly 
with $\nc$, just like the action
of a continuum instanton.
But we see that there are configurations with non-zero $Q$, and the question 
of their
survival at infinite $\nc$ is similar to the familiar question in the continuum 
\cite{bridgehead}.
This indicates that in four dimensions,
the reduced lattice model could provide a framework in which one can at least
pose questions about topology at infinite $\nc$. This is interesting because
the results of Teper \cite{teper} support a finite nonvanishing limit to the topological
susceptibility in the lattice planar limit. 
We do not know if there are  
configurations that produce $Q=1$ and have an action
that stays finite as $\nc$ goes to infinity.

We now make an observation that might be of relevance to the issue 
of $\theta$-dependence in the planar limit. The set of backgrounds
satisfying (\ref{rest}) falls apart into non-empty disconnected
components beyond what is needed for distinct topology.
Even at fixed $Q$ there are several components. 
These different components make separate contributions
to the partition function, and these contributions are not obviously suppressed
at infinite $\nc$. It could be that the particular connected
component that dominates at infinite $\nc$ depends 
discontinuously on parameters, and therefore the action density
need no longer be differentiable with respect to its parameters.
In particular, this could induce a nonanalyticity in 
$\theta/\nc$, and this
might be a way of reconciling the naive scaling 
for the action density ($\sim\nc^2 f (\frac{\theta}{\nc} )$) 
with $2\pi$ periodicity in $\theta$ ~\cite{witten}.

To see that the set of gauge fields one integrates
over splits into disconnected components, consider again the
above two dimensional example. Complete the pair $T_{1,2}$ by
a $T_3$ and a $T_4$ of a perturbative nature so that $Q=0$. 
The shift-clock pair cannot be deformed smoothly to unity while obeying the
constraint on the commutator because the shift-clock pair could be used to construct
a two dimensional lattice Dirac operator which would have a two dimensional
topological charge of unity. 
Thus, this set of four $T_\mu$'s has $Q=0$ but is homotopically inequivalent
to a set where all four $T_\mu$'s are perturbative and consequently  
also has $Q=0$. 
In spite of the fact that we are considering
only $SU(\nc )$ and not $U(\nc )$, the two dimensional Chern numbers associated with
the different two dimensional tori included in the four dimensional torus 
manage to survive in the reduced model.

Let us briefly mention anomalies associated with chiral fermions. 
The free energy of the reduced model with fermions included is
supposed to reproduce the free energy of the original model at
leading order $\nc^2$ and subleading order $\nc$. Hence, anomalies
should spoil the gauge invariance of the reduced model too. 
In the overlap formalism, there are simple lattice abelian gauge backgrounds \cite{geom}
in which one can prove that gauge invariance cannot be preserved if there
are anomalies in the continuum. Moreover, this happens already on finite
lattices. These backgrounds can be ``borrowed'' to produce nonabelian backgrounds
in the reduced model with a similar effect. 
The overlap line bundle is twisted over the torus defined by 
variables $|p_{1,2}|\le \pi $ that enter through the 
family of gauge backgrounds described by 
$T_1 =e^{ip_1}\otimes Y$, $T_2=e^{ip_2}\otimes Y$, $T_3=X\otimes 1$,  
$T_4=Y\otimes 1$ and, as in the case of the ``instanton'', $\nc=N^2$. 
Again, we see the potential to move 
beyond the obstacles identified in \cite{grosskita}, but there is a degeneracy
of eigenvalues we need to eliminate.

\subsection{Fermions}

In the original unreduced lattice model, massless fermions are incorporated
through the massless overlap Dirac operator \cite{overlap}
\begin{equation}
         D_o = \frac{1}{2} ( 1 + \gamma_{d+1} \epsilon )
\end{equation}
with $\epsilon = {\rm sign}(H_W)$. The operator $D_o$ has an implicit 
dependence on the fermion parallel transporter. 

The reduced version of $D_o$ is obtained by replacing the original transporter
with the reduced fermion parallel transporter. For fermion momenta
$-\pi < p_{\mu} \leq \pi$  and $T_{\mu}$ given in (\ref{tmured}), 
the reduced transporter
is $e^{i p_{\mu}} T_{\mu}$. With that replacement, $D_o$ becomes the reduced
massless overlap Dirac operator. It depends on the fermion momentum. The
external line propagator~\cite{int-ext} is 
\begin{equation}
   g(p) = \frac{1}{2m} [ D_o^{-1} -1 ]
\end{equation}

Consider the connected correlation function of the fermion bilinear $\frac{1}{\sqrt{\nc}}\bar\psi (x) \psi (x)$. In momentum space it
is given by
\begin{equation}
\label{mesonprop}
G(P) = -\frac{1}{\nc} \langle [\langle Tr g(p) g(q) \rangle_{\rm gauge~average}]
\rangle_{\rm angle~average}
\end{equation}
with $P=p-q$.  
Similar expressions can be derived for the correlation functions
of other bilinears. 

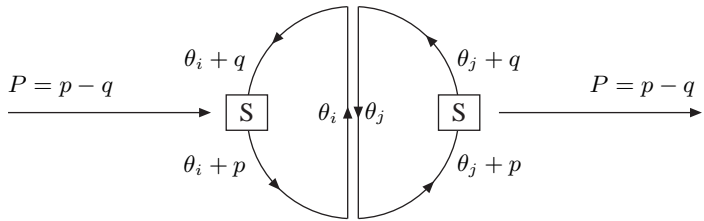
\begin{figure}

\begin{center}
\begin{picture}(300,100)(0,0)
\ArrowArc(150,50)(40,-87,0)
\ArrowArc(150,50)(40,0,87)
\ArrowLine(148,10)(148,90)
\ArrowLine(152,90)(152,10)
\ArrowArc(150,50)(40,93,180)
\ArrowArc(150,50)(40,180,267)
\BText(110,50){S}
\BText(190,50){S}
\Text(110,70)[r]{\small $\theta_i +q$}
\Text(110,30)[r]{\small $\theta_i +p$}
\Text(190,70)[l]{\small $\theta_j +q$}
\Text(190,30)[l]{\small $\theta_j +p$}
\Text(145,50)[r]{\small $\theta_i$}
\Text(155,50)[l]{\small $\theta_j$}
\LongArrow(20,50)(95,50)
\LongArrow(205,50)(280,50)
\Text(20,60)[l]{$P=p-q$}
\Text(280,60)[r]{$P=p-q$}

\end{picture}\\
\end{center}
\caption{A planar diagram representing the one gluon exchange
contribution to the scalar - scalar correlation at momentum $P$.}
\label{mesprop}
\end{figure}

The way reduction works is best understood by looking at an example.
Figure~\ref{mesprop} is a diagram showing the various momenta in the one gluon exchange contribution
to a scalar-scalar correlator.
All crystal momenta come from the diagonal entries of the fermion transporter
and flow in the directions indicated by the arrows. The $\theta$ momenta
are associated with color and the $p,q$ momenta with flavor. Single lines are
fermions, and the double line is the exchanged gluon. The gluon carries momentum
$\theta_i -\theta_j$. With the quenched $\theta_i$ randomly distributed on the
momenta tori, the summation over the color indices $i$ and $j$ induces
the needed integrations over momenta. 

Note that we compute the meson correlators directly in momentum space. 
This is an important
new feature with several potential applications. 

\vskip .2in

When the quarks are massive, the propagator $g(p)$ is \cite{urs}
\begin{equation}
\label{massprop}
g(p)=\frac{1}{2m} 
       \frac{1-\gamma_{d+1} \epsilon}{(1+\mu) +(1-\mu) \gamma_{d+1} \epsilon}.
\end{equation}
The parameter $\mu$ fixes the bare lattice quark mass via $m_q = 2m \mu$. 

The transporter $e^{i p_{\mu}} T_{\mu}$ introduces a phase into the reduced
fermion operator. 
This suggests that there are several options for defining $Q$,
something we glossed over in the previous subsection. It is a bit strange
that one seems to be able to define a topological charge that depends
on the momentum carried by the fermions. Probably, for
configurations typical to the continuum limit,
$Q$ ends up not depending on $p$. This is indeed the case for examples
in section IIB as long as $N$ is large enough and the momenta are not
too large.

It is quite easy to add $N_f$ flavors of fermions
and then go to the topological limit $N_f, N_c \to\infty$ with
$N_f/N_c$ kept fixed. In this limit, the fermionic determinant needs to be
included in the gauge field integration.
Similarly, one can also introduce fermions in representations other
the fundamentals.
Chiral gauge theories will have anomalies because, as mentioned before, the 
backgrounds that produce them
have a reduced model representative~\cite{geom}. 

\section{Two dimensional planar QCD}
As an exercise, we will calculate the scalar-scalar 
\be
\label{sc}
\langle \bar\psi (x)\chi(x)\bar\chi(0) \psi (0)\rangle 
\ee
and the pseudoscalar-pseudoscalar
\be 
\label{psc}
\langle \bar\psi (x)\gamma_3 \chi(x)\bar\chi(x)\gamma_3 \psi (0)\rangle 
\ee 
correlation functions in momentum space in the planar limit
of two-dimensional QCD. (We use two different fields,
$\psi$ and $\chi$ to eliminate the disconnected piece.) 
These correlation functions are known
exactly and are completely determined by their Feynman diagrams. Nevertheless,
they are quite nontrivial. While they can be
expressed as an infinite sum over stable mesons with mass squares that are
evenly spaced asymptotically, they also have free field behavior at
large momenta.  Thus, we have the 
cohabitation of asymptotic freedom and total confinement
that is also present in four-dimensional planar QCD.
Moreover, with an appropriate choice of limits, chiral symmetry is broken,
and there a Goldstone boson.
If the potential infrared problems turn out to be tractable in two dimensions,
we can be confident that there will not be any difficulties in four 
dimensions where infrared issues are less serious.

In two-dimensional QCD,  
spontaneous chiral symmetry breaking only occurs if one takes 
the limit $m_q\to 0$ after the planar limit $\nc\to\infty$ at fixed 
$\beta/\nc$.
In the reduced model for any finite $\nc$, we have exact chiral symmetry
at $m_q=0$, and we can investigate the planar limit of the massless
reduced model. The continuum limit is taken last.
With $m_q$ set to zero before the large $N_c$ limit, we will 
not see direct signals of spontaneous symmetry breaking.
The simplest assumption is that chirally invariant
quantities will be well-described by 't Hooft's solution,
but chirally noninvariant observables will not be.
Thus the invariant combination to compare to the 't Hooft solution is  
\be 
\label{sspp}
\langle\bar\psi(x)\chi(x)\bar\chi(0)\psi(0)
\rangle - \langle\bar\psi(x)\gamma_3 \chi(x)\bar\chi(0)\gamma_3 \psi(0)
\rangle .
\ee

For massive quarks, the situation is less subtle, and one can consider the
correlators separately.
The planar limit of the reduced model
is supposed to produce a UV regulated approximation to the
't Hooft solution. It is interesting to see in detail how this works.
On the one hand, the very high mass mesons are meaningless entities at
finite ultraviolet cutoff, while on the other hand, they are responsible for
the free field short distance behavior.  

It is believed that ordinary, massive two-dimensional QCD has two phases 
\cite{QCD2a}: a weak
coupling one in which the gauge couping divided by the quark mass is
small and a strong coupling one in which this ratio is large. In the
strong coupling phase, there are light baryonic states,
but there is no pion
whose mass vanishes as $\sqrt{m_q}$ in the chiral limit, 
and it is not clear
how the large $\nc$ limit should be defined. 't Hooft's planar solution is in 
the weak coupling phase
and does have a pion mass that vanishes as $\sqrt{m_q}$.
It is interesting that the numerical results obtained
in \cite{federico} seem to correspond to the strong coupling phase
and, nevertheless, also seem to admit a large $\nc$ limit with the
't Hooft scaling of the gauge coupling. Thus it could be that in 
two-dimensional QCD, there exists a strong coupling planar 
phase that is defined through the lattice and which differs from 
't Hooft's solution. 

We will work in the weak coupling phase where 't Hooft's solution
should be valid. If there are indeed two planar phases, convergence
to the planar continuum limit of the weak coupling phase 
might be slower than anticipated.

\subsection{Pure gauge}

We can go to a ``unitary gauge'' in which $T_2$ is fixed to
the clock matrix $Y$, with $N=\nc$. Denoting $T_1$ by $U$ the
partition function is
\begin{equation}
Z=\int_{U=W Y W^\dagger,~~W\in U(N)} dW e^S~~~~~~~
~~S=4\beta \sum_{ij} |U_{ij}|^2 \sin^2 \frac{\pi (i-j)}{\nc} .
\end{equation}
Terms with $i=j$ do not contribute to $S$. All absolute minima of $S$
are at $U^\sigma_{ij} = \delta_{ij} e^{\frac{2\imath\pi \sigma_i}{\nc}}$. 
 Here $\sigma$ is a permutation of $0,1,2,\dots, \nc -1$. There are no other minima.

For very large $\beta/\nc$, the relative weights of distinct minima are given 
by the
ratios of determinants of small fluctuations around the configurations.
We parametrize the fluctuating degrees of freedom by the off diagonal entries
of a hermitian matrix $H$ which generates a conjugation around a diagonal saddle.
\begin{equation}
\delta U_{ij} = i[U^\sigma,H]_{ij} = iH_{ij} [e^{\frac{2\imath\pi \sigma_j}{\nc}}-
e^{\frac{2\imath\pi \sigma_i}{\nc}}]
\end{equation}
The diagonal terms of $U$ don't contribute, and the second order action is
\begin{equation}
\delta^2 S = 16\beta\sum_{ij} |H_{ij}|^2 \sin^2 \frac{\pi(\sigma_i-\sigma_j)}{\nc}
\sin^2\frac{\pi(i-j)}{\nc}  .
\end{equation}
Since
\begin{equation}
\prod_{i\ne j} |\sin \frac{\pi(\sigma_i-\sigma_j)}{\nc}|=
\prod_{i\ne j} |\sin \frac{\pi(i-j)}{\nc}|
\end{equation}
for any permutation $\sigma$, 
all minima should be visited with equal frequency for large $\beta/\nc$.

One may ask whether the barriers between different
minima are high and whether they become infinite in the large $\nc$ limit.
The minima can be connected by sequences of transpositions. We do not know 
whether the barriers are infinite at infinite $\nc$ and whether in two
dimensions, where the Feynman
diagrams capture all the physics, one can ignore the many minima.

From now on, we will use $b=\sqrt{\frac{\beta}{\nc}}$ as the parameter held
fixed as $\nc$ goes to infinity. In the planar limit, one approaches the 
continuum limit
(and hopefully the 't Hooft solution) when $b\to\infty$. Comparing to a 
continuum
definition of the gauge coupling where the partition function is given by
\begin{equation}
\exp\{-\frac{1}{4g^2} \sum_{\mu\nu} Tr F_{\mu\nu}^2\}  ,
\end{equation}
we get 
\begin{equation}
g^2=\frac{1}{2\beta}=\frac{1}{2\nc b^2}     .
\end{equation}

To get a feeling for the numbers, consider two-dimensional QCD with massless 
quarks.
The lowest scalar meson mass is approximately $0.967/b$.
In taking a sequence of increasing $\beta$'s and $\nc$'s
with $b^2=5$, we find that this mass approaches $0.43$ in lattice units.
With a mass scale this large in lattice units, we can expect good agreement
with the continuum theory only for small momenta.

In two dimensions, quenching is not necessary \cite{bhn,gold}. We will use a 
version
of the model where there is no longer a need to perform the average
over the angles. In this version, the eigenvalues of the $T_\mu$ matrices
are still frozen to the set consisting of the $\nc$ roots of unity.
This adds a new ``translational" symmetry to 
the model in which each $T_\mu$ is replaced by
$z_\mu T_\mu$ with $z_\mu^{\nc}=1$. This transformation preserves the eigenvalue
sets. With $\nc$ chosen to be a prime number, the eigenvalues of $T_\mu^k$ will
span the same roots of unity 
for any $0\le k\le \nc-1$, and will simulate the situation on a torus of size
$\nc^2$. 

Let us summarize the symmetries of the bosonic action:
\bea
\label{symm}
&S_g (T_1 , T_2)= S_g (z_1 T_1, z_2 T_2)~~{\rm with}~~|z_\mu|=1\nonumber\\
&S_g (T_1 , T_2)= S_g (T_1^\dagger , T_2)=S_g (T_1, T_2^\dagger)\nonumber\\
&S_g (T_1, T_2) = S_g (T_2 , T_1)
\eea
These symmetries reflect invariances of the original toroidal lattice theory.

\subsection{Fermions and meson observables}

The fermionic momenta $p_\mu$ can be chosen from the set $\frac{2\pi j}{\nc}$,
$j=0,1,\dots,\nc-1$ for fermions with periodic boundary conditions
and from the set $j=\frac{1}{2},\frac{3}{2},\dots,\nc-\frac{1}{2}$
for fermions with anti-periodic boundary conditions. Thus, the momenta
flowing in the fermion lines are identical in type
to those flowing
in the gauge boson lines. With these momenta, the planar Feynman diagrams are 
similar to
those of a system on a lattice with $\nc^2$ sites. 

The symmetries of (\ref{symm}) imply that the meson field propagators (\ref{mesonprop})
obey the following relations:
\bea
G(P_1 ,P_2 )= G(-P_1 ,P_2 )=G(P_1 ,-P_2 )=G(-P_1 ,-P_2 )\nonumber\\
G(P_2 ,P_1 )=G(P_2 , -P_1 )=G(-P_2 ,P_1 )=G(-P_2 ,-P_1 )
\eea
These symmetries will be strictly enforced by explicitly adding the 
contributions
of all gauge field configurations connected by the symmetry transformations
of (\ref{symm}). 

$G$ will be denoted by $S$ for the scalar meson propagator and by $P$ in the
pseudoscalar case. 't Hooft's solution, together with 
subsequent work \cite{ccg}, provide
explicit formulae for $S$ and $P$ in momentum space. Since they are
composite, there are ultraviolet divergences, and the unrenormalized
formulae are divergent.
Since their source is in free field theory, these divergences can be regulated 
in many
ways. We define the renormalized
meson correlators $S_R$ and $P_R$ by subtracting at zero momentum for massive quarks and
at an arbitrary momentum point for massless quarks. 

The continuum results we strive to reproduce using our reduced model are
expressed in terms of the eigenvalues and eigenfunctions of 't Hooft's Hamiltonian
$H$. It acts on square integrable functions defined on the
interval $[0,1]$. The boundary conditions are Dirichlet for massive quarks
and von Neumann for massless quarks. For massive quarks, $H$ is positive 
definite.
For massless quarks, it has one eigenstate with zero eigenvalue, and all other
eigenvalues are strictly positive. There are no degeneracies, and the eigenvalues
of $H$ provide the masses squared of the meson bound states. In the following
discussion, all dimensionful
quantities are converted to their dimensionless counterparts using the scale
\be
e^2\equiv\frac{g^2\nc}{\pi}
\ee
which has the units of mass squared.
In these dimensionless variables, 
\be
H \phi_n = \mu_n^2 \phi_n,~~~~n=0,1,2,\dots  .
\ee
The quark mass enters by the dimensionless parameter 
\be
\gamma=\frac{m_q^2}{e^2}   .
\ee
The squared momentum $P^2$ is replaced by the dimensionless variable $Q^2$
\be
Q^2=\frac{P^2}{e^2}   .
\ee

The basic formulae for the meson propagators with massive quarks are:
\bea
\label{thform}
&P_R (Q^2, \gamma ) = \sum_{n\ge 0~{\rm even}}^\infty \left [ 
\frac{r_n^2}{Q^2 +\mu_n^2} - \frac {r_n^2}{\mu_n^2}\right ] \nonumber\\
&S_R (Q^2, \gamma ) = \sum_{n\ge 1~{\rm odd}}^\infty \left [ 
\frac{r_n^2}{Q^2 +\mu_n^2} - \frac{r_n^2}{\mu_n^2} \right ]
\eea
Asymptotically $\mu_n^2 \sim \pi^2 n$, and the residues approach a finite
limit, so the subtracted sums converge.

The residues $r_n^2$
are given in terms of the real, normalized wavefunctions $\phi_n (x)$:
\be
r_n^2 =\frac{1}{\pi} \gamma \left [ \int_0^1 dx\frac{\phi_n (x)}{x} \right ]^2
\ee

For massless quarks, one needs to take the limit $\gamma\to 0$. 
The relevant combination is 
$S_R+P_R$, which is the regulated version of (\ref{sspp}). 
For $m_q=0$, the lowest meson mass is zero, and there is an infrared
divergence at $p=0$. This is handled by making the subtraction at 
nonzero $Q^2$.
More information about $H$ and the evaluation of  (\ref{thform}) is collected
in the appendix. 

\subsection{Numerical results in the reduced model.}

We have focused on two cases. The first is $\gamma=1$ which,
as explained in the Appendix, corresponds to an intermediate quark mass.
The other case is massless quarks, where
the chiral symmetry is spontaneously broken at weak $g^2$-coupling in 
the planar limit. Other values of $\gamma$ are briefly discussed.

Numerical results are obtained by working in the ``unitary gauge''
and generating gauge fields $T_1$ that are related to $Y$ by conjugation.
Every conjugation belongs to an $SU(2)$ subgroup of $SU(\nc)$.
A proposed conjugation
is accepted or rejected by the Metropolis algorithm. The $SU(2)$ elements are
picked so that the acceptance rate 
is close to 0.5. One iteration is defined as a sequence of attempted updates
for an entire set of ${\nc(\nc-1)\over 2}$ $SU(2)$ subgroups of $SU(\nc)$. The 
subgroups in the set are defined by placing
the $SU(2)$ matrix in the $(n,n),(n,m),(m,n),(m,m)$ entries of 
the $SU(\nc)$ $T_1$ matrix. The indices obey $ 1 \le n < m \le \nc$.
Fermionic measurements are made every 100 iterations, and 1000 iterations
are used for thermalization.

Fermionic propagators are computed using (\ref{massprop}). This
is done by diagonalizing $H_W$ exactly, constructing $\epsilon$,
and then performing an exact diagonalization of $\gamma_3 \epsilon$. 
The propagator,
$g(p_1,p_2)$,
at any mass, is obtained from the spectral decomposition of
$\gamma_3 \epsilon$. 
This is done for all values of $p_1$ and $p_2$ and at a fixed value
of $T_1$. The ``bare'' scalar and pseudoscalar propagators are 
\begin{eqnarray}
S_0 (P) &=& {1\over 4N_c^2} \sum_{p_1,p_2} 
Tr [ g(p_1+P_1,p_2+P_2) g(p_1,p_2)\cr
&+& g(p_1-P_1,p_2+P_2) g(p_1,p_2) \cr
&+& g(p_1+P_2,p_2+P_1) g(p_1,p_2) \cr
&+& g(p_1+P_2,p_2-P_1) g(p_1,p_2) ]
\end{eqnarray}
\begin{eqnarray}
P_0 (P) &=& {1\over 4N_c^2} \sum_{p_1,p_2} 
Tr [ g(p_1+P_1,p_2+P_2) g^\dagger(p_1,p_2) \cr
&+& g(p_1-P_1,p_2+P_2) g^\dagger(p_1,p_2) \cr
&+& g(p_1+P_2,p_2+P_1) g^\dagger(p_1,p_2)\cr 
&+& g(p_1+P_2,p_2-P_1) g^\dagger(p_1,p_2) ] .
\end{eqnarray}
There are different vectors $P=(P_1,P_2)$ which correspond
to the same value of $P^2\equiv 4\sum_\mu \sin^2{P_\mu\over 2}$.
By definition, the lattice meson propagators are a function of
$P^2$ only. This function is obtained
by averaging over all momenta with the same $P^2$. 
After a final average over several values of
$T_1$, these propagators are the reduced model's approximation to (\ref{thform}).

\begin{figure}
\epsfxsize = 0.8\textwidth
\centerline{\epsfbox{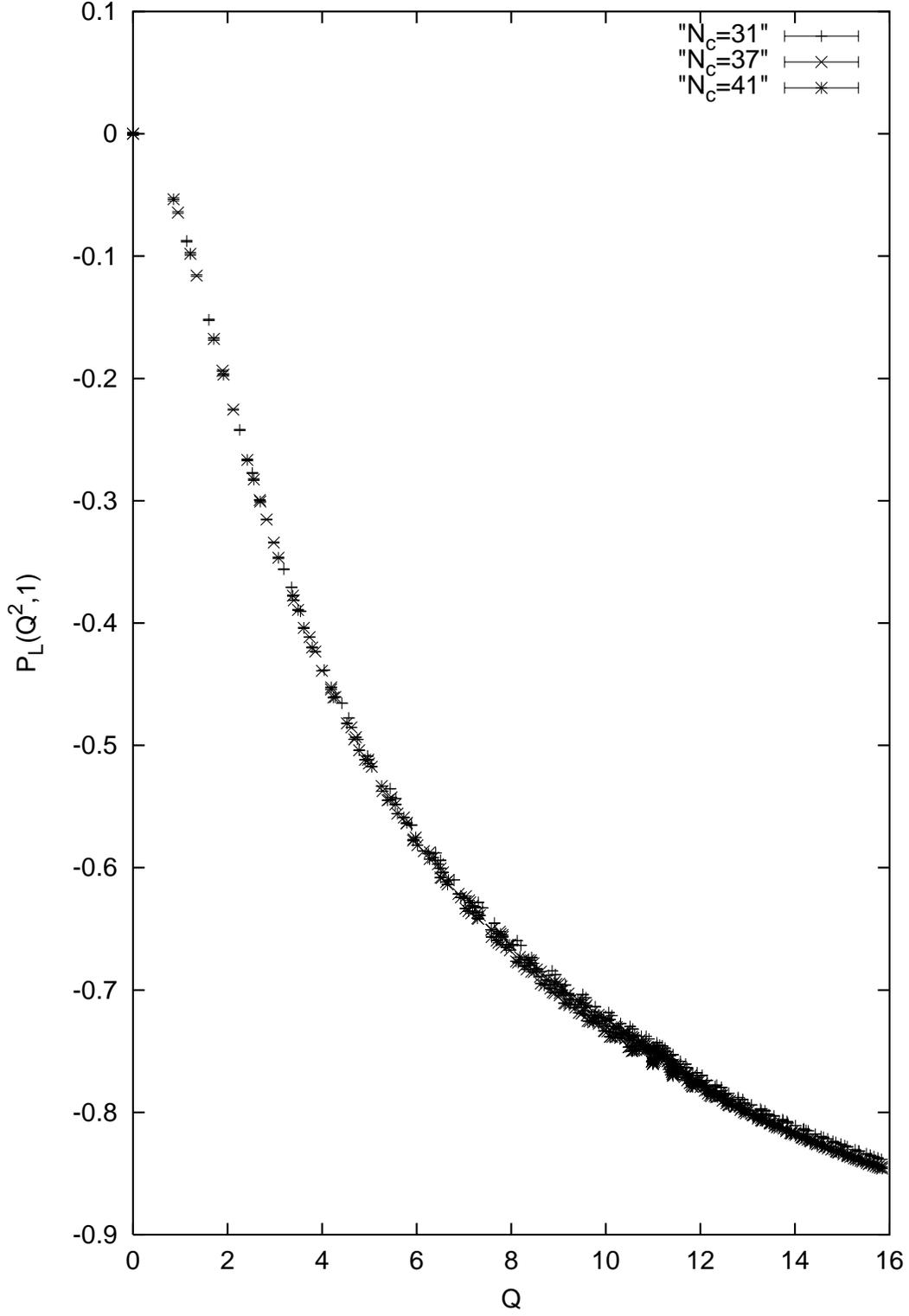}}
\caption{
Plot of $P_L(Q^2,1)$ at $b^2=5$ for $N_c=31,37,41$.
}
\label{b25Pg1}
\end{figure}

\begin{figure}
\epsfxsize = 0.8\textwidth
\centerline{\epsfbox{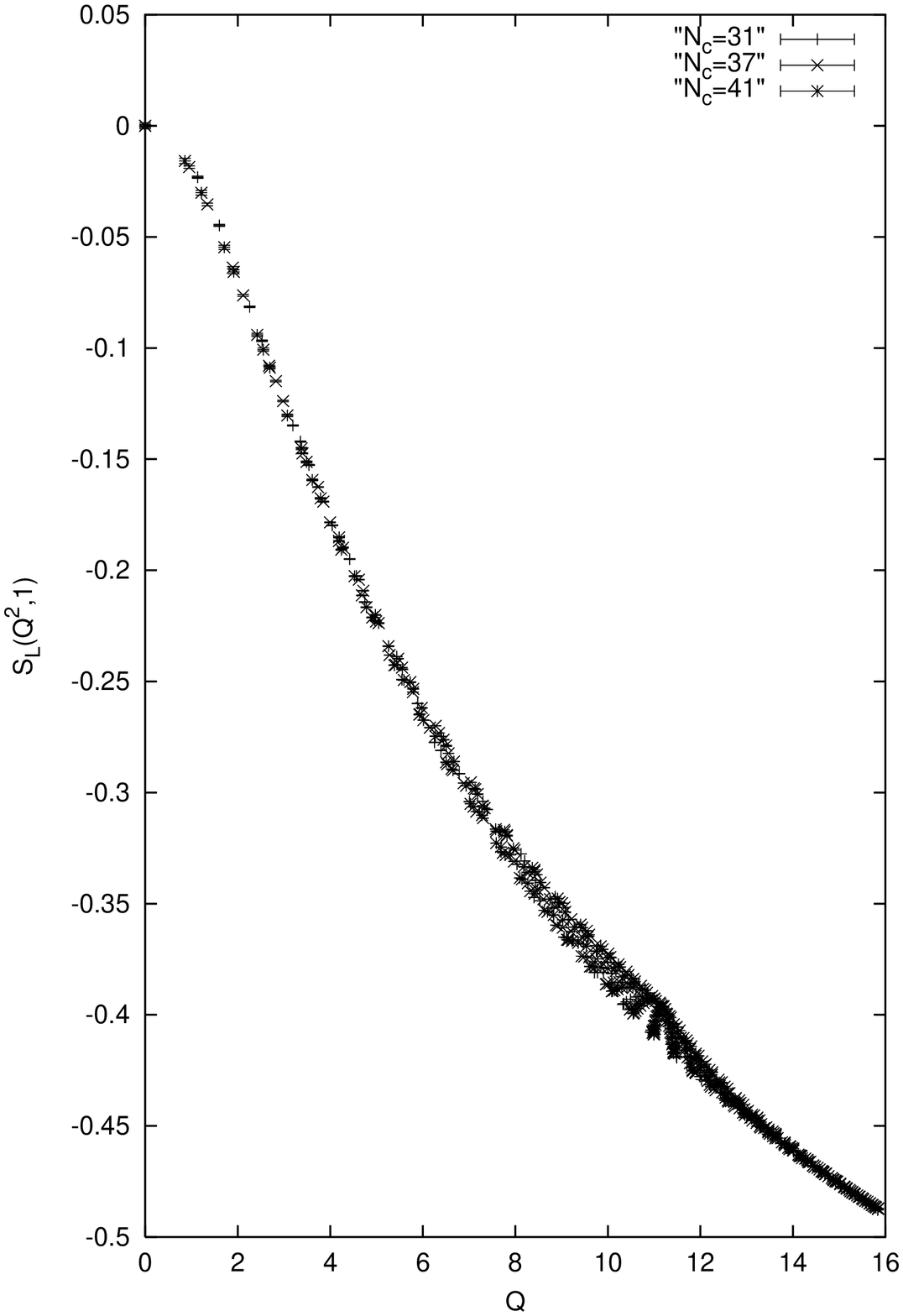}}
\caption{
Plot of $S_L(Q^2,1)$ at $b^2=5$ for $N_c=31,37,41$. 
}
\label{b25Sg1}
\end{figure}

\begin{figure}
\epsfxsize = 0.8\textwidth
\centerline{\epsfbox{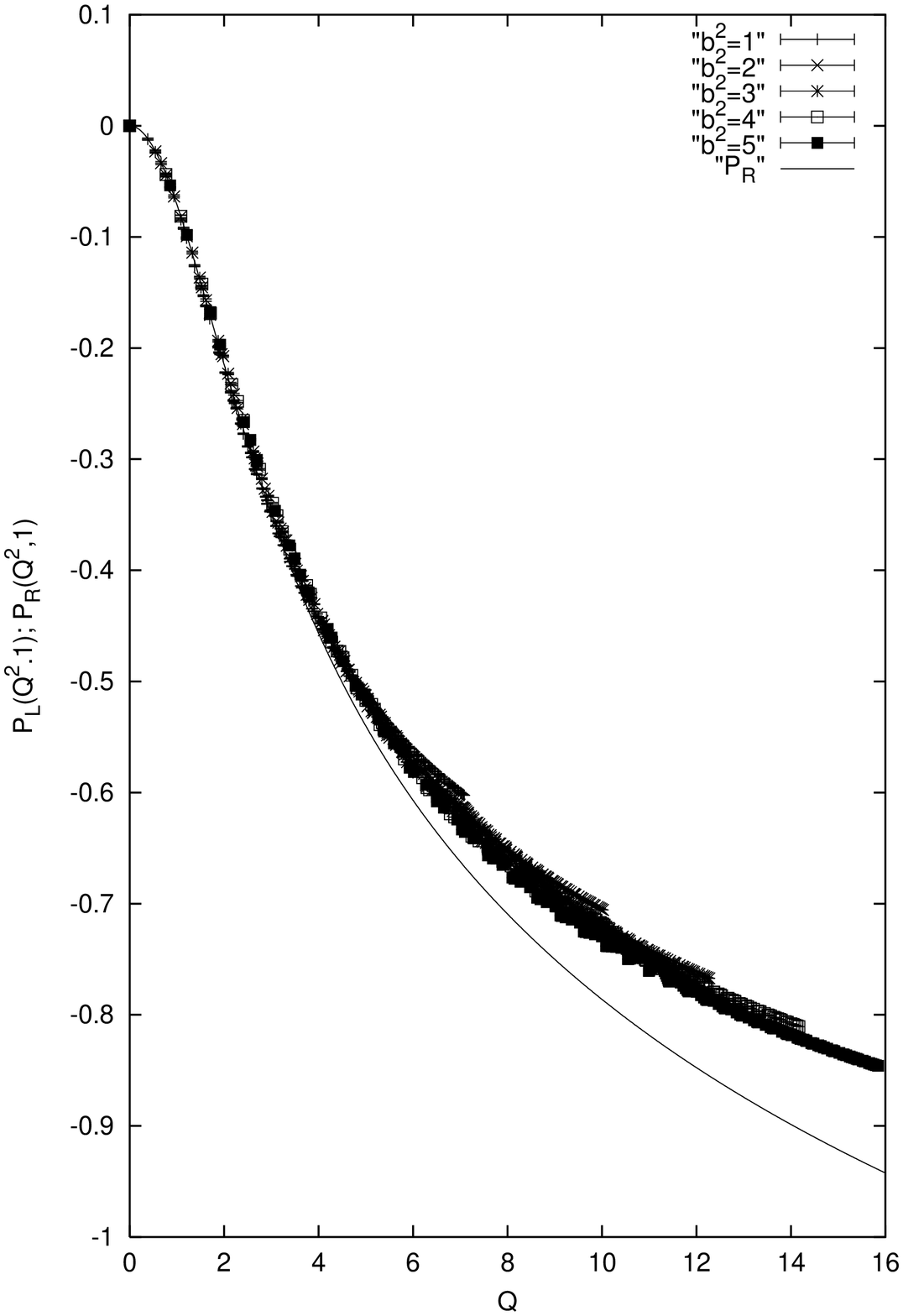}}
\caption{
Plot of $P_L(Q^2,1)$ at $N_c=41$ for $b^2=1,2,3,4,5$. 
}
\label{pseudog1}
\end{figure}

\begin{figure}
\epsfxsize = 0.8\textwidth
\centerline{\epsfbox{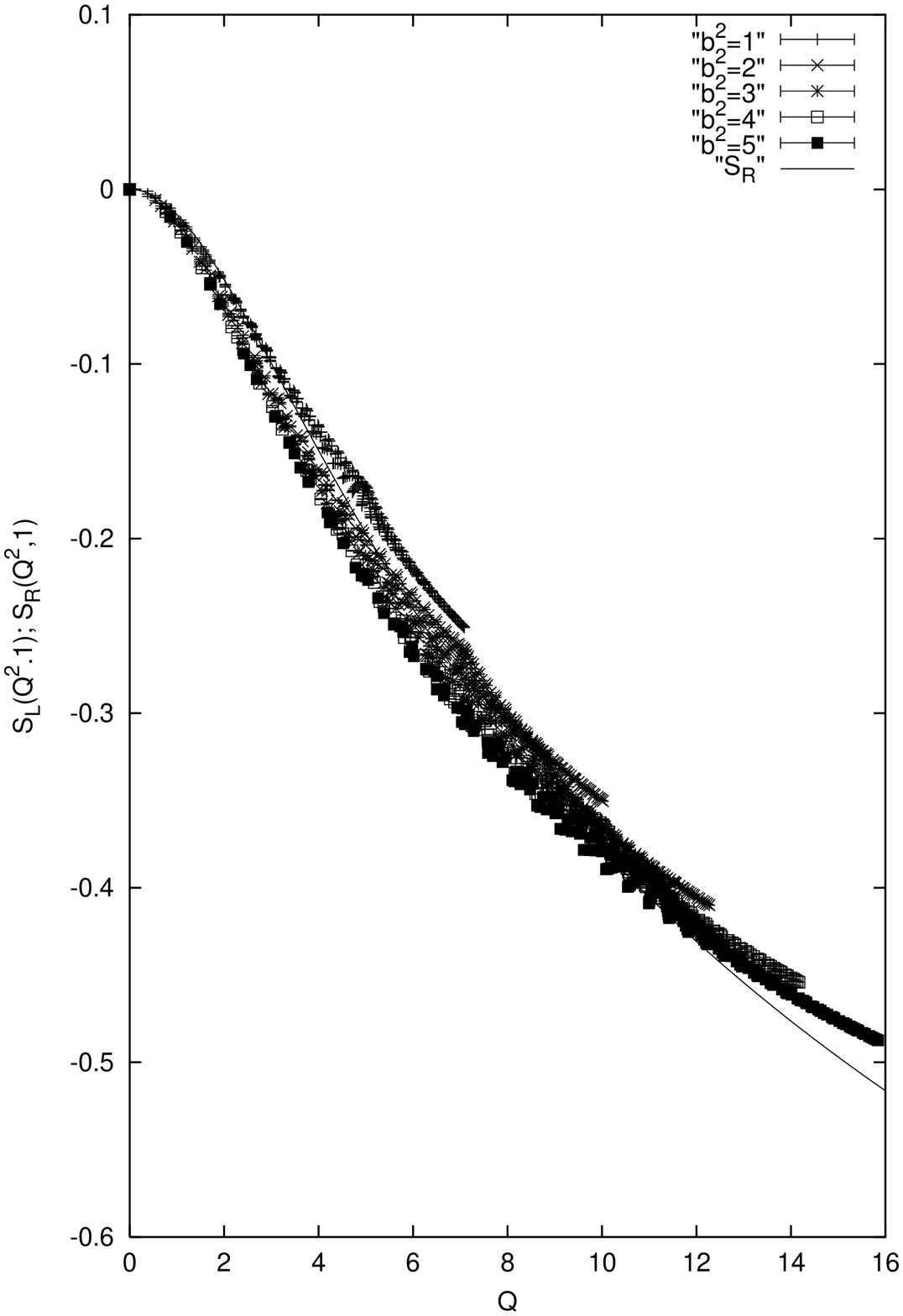}}
\caption{
Plot of $S_L(Q^2,1)$ at $N_c=41$ for $b^2=1,2,3,4,5$. 
}
\label{scalarg1}
\end{figure}

We first compare numerical results to the continuum keeping 
$\gamma=1$ fixed. 
In order to study the approach to the continuum limit,
we carried out simulations at five values of $b^2$: $1,2,3,4,5$.
At each value of $b^2$, we generated gauge field configurations at
$\nc =31, 37,$ and $41$ to see how the limit $\nc=\infty$ is approached.  
The lattice momenta $\sqrt{P^2}$ range from $0$ to $2\sqrt{2}$
while $\sqrt{Q^2}$, reaches a maximum of $7$ at $b^2=1$ 
and a maximum of $16$ at $b^2=5$. In this range of momenta, 
good estimates of $P_R(Q^2,1)$ and $S_R(Q^2,1)$ are obtained by 
including only the first $100$ terms in the sum over states. 
Since we have included 
the first $2000$ eigenvalues (see Appendix), 
our ``theoretical'' data are very accurate.

Let $S_L(Q^2,1)$ and $P_L(Q^2,1)$ denote the regularized, reduced model scalar
and pseudoscalar propagators. As in
the continuum, we subtract the propagators at zero momentum. 
Data at $b^2=5$ are shown in
Figs. \ref{b25Pg1} and \ref{b25Sg1}. At lower values of $b$,
the $N_c$ dependence is even weaker.
Even at $b^2=5$, we find that 
$\nc=41$ is large enough to provide accurate numbers for $\nc=\infty$.

We study the approach to the continuum limit in 
Figs. \ref{pseudog1} and \ref{scalarg1}. Looking 
at different $b^2$ values, we see that the pseudoscalar propagator 
approaches its continuum limit more smoothly than the scalar
propagator. For $\sqrt{Q^2} < 4$, the numbers seem to have converged
to their $b^2 \to\infty$ limit, and there is reasonable
agreement with the continuum result. 
On the other hand, the scalar propagator
has not yet converged to its $b^2=\infty$
limit in any region of $Q^2$. It shows some overall agreement with continuum
results, but it is less convincing than in the pseudoscalar case. In addition, 
the pseudoscalar propagator seems to approach its continuum limit
monotonically, but the scalar propagator does not.

\begin{figure}
\epsfxsize = 0.8\textwidth
\centerline{\epsfbox{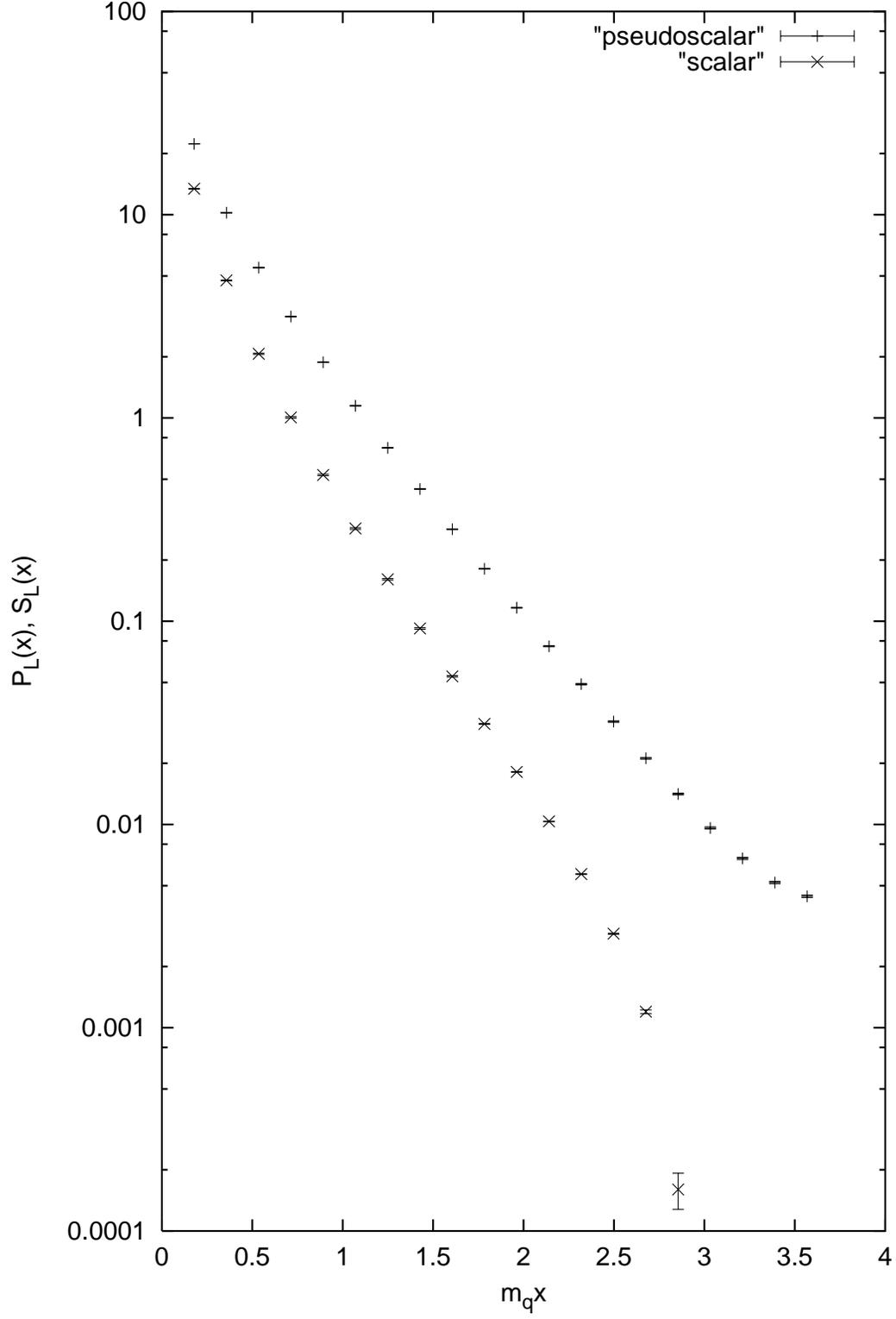}}
\caption{
Plot of the scalar and pseudoscalar propagators in real space at
$b^2=5$ and $N_c=41$.
}
\label{proprs}
\end{figure}

The main difference between the scalar and pseudoscalar propagators
is that the lowest mass in the pseudoscalar channel is smaller
than the lowest mass in the scalar channel. In the pseudoscalar propagator,
we have agreement below $\sqrt{Q^2}=4$, and 
the lowest pseudoscalar mass is $\mu_0=2.69713$, which is below $\sqrt{Q^2}=4$. 
However, the lowest scalar mass is $\mu_1=4.16036$, which is slightly above
$\sqrt{Q^2}=4$. 
Hence, all the masses contributing to the scalar propagator 
are above the region
of $Q$ where the pseudoscalar propagator converged well to its continuum
limit. Thus, the difference we found between the scalar and pseudoscalar
propagators is likely to be a finite $b^2$ effect. 

\begin{figure}
\epsfxsize = 0.8\textwidth
\centerline{\epsfbox{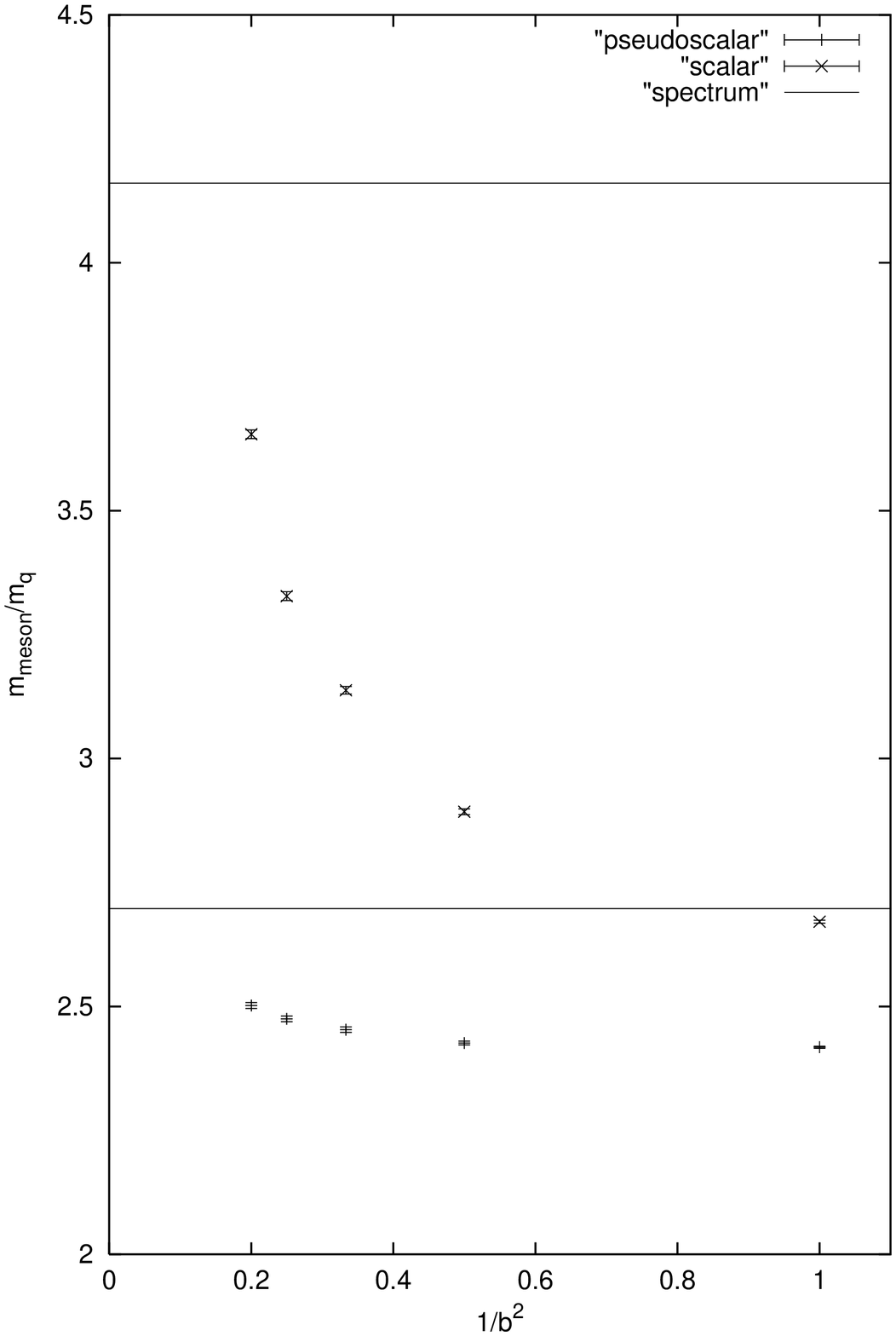}}
\caption{
Plot of the effective mass extracted from the scalar and pseudoscalar
propagators as a function of
$b^2$ at $N_c=41$.
}
\label{massfit}
\end{figure}

The difference between the two propagators can also be seen by
extracting an effective mass. We set the momentum in one direction
to zero and Fourier transform on the other component to produce
a function of a conjugate, discrete ``spatial'' variable. 
Figure \ref{proprs} is a semilog plot of the 
scalar and pseudoscalar propagators at sample parameter values $b^2=5$
and $\nc=41$.
We see that the pseudoscalar propagator
is a better fit to a straight line. 

One can
get an effective mass in either channel
by forcing straight line fits.  
The resulting effective masses at $\nc=41$ are shown
in Fig. \ref{massfit} as a function of $b^2$. 
Although both masses appear to approach the
correct continuum values, the finite $b^2$ effects on 
the scalar mass are much stronger. 

\begin{figure}
\epsfxsize = 0.8\textwidth
\centerline{\epsfbox{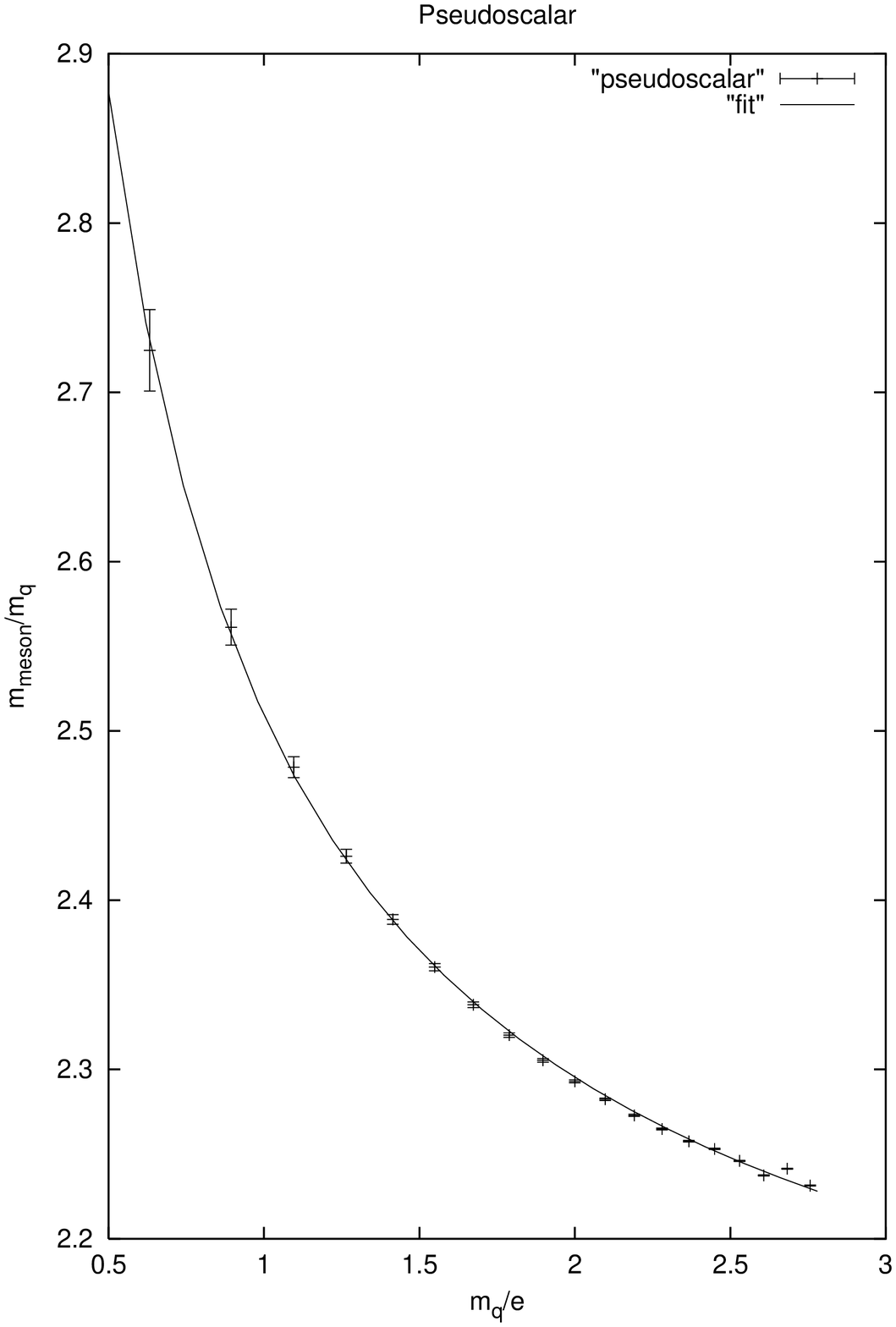}}
\caption{
Plot of the effective mass extracted from the pseudoscalar
propagator as a function of
$m_q/e$ at $N_c=37$.
}
\label{massfg}
\end{figure}

We now turn our attention to the behavior of the pseudoscalar propagator
as a function of $m_q/e$. We numerically computed this propagator as
a function of $\sqrt{P^2}$ at several
values of $m_q/e$ for $b^2=5$ and $\nc=37$. 
An effective mass was extracted as before. This effective mass
is plotted as a function of $m_q/e$ in Fig. \ref{massfg}. 

The empirical fit 
\be
\frac{m_{\pi}}{m_q} = 2 +0.51 ({m_q\over e})^{-0.78}
\ee
shown in the plot works quite well. 

With increasing $m_q/e$, the ratio $m_\pi/m_q$ approaches $2$,
as expected. However, the subleading behavior does not match the expression
derived in the Appendix which has $-1.33$ rather than $-0.78$
in the exponent. 
On the other hand, $m_\pi/e$ is clearly not proportional
to $\sqrt{m_q/e}$ as would be appropriate for light quarks.
We get $0.78$ instead of $0.5$ in the exponent.
The fit indicates that the masses we used are in an intermediate
region.

\begin{figure}
\epsfxsize = 0.8\textwidth
\centerline{\epsfbox{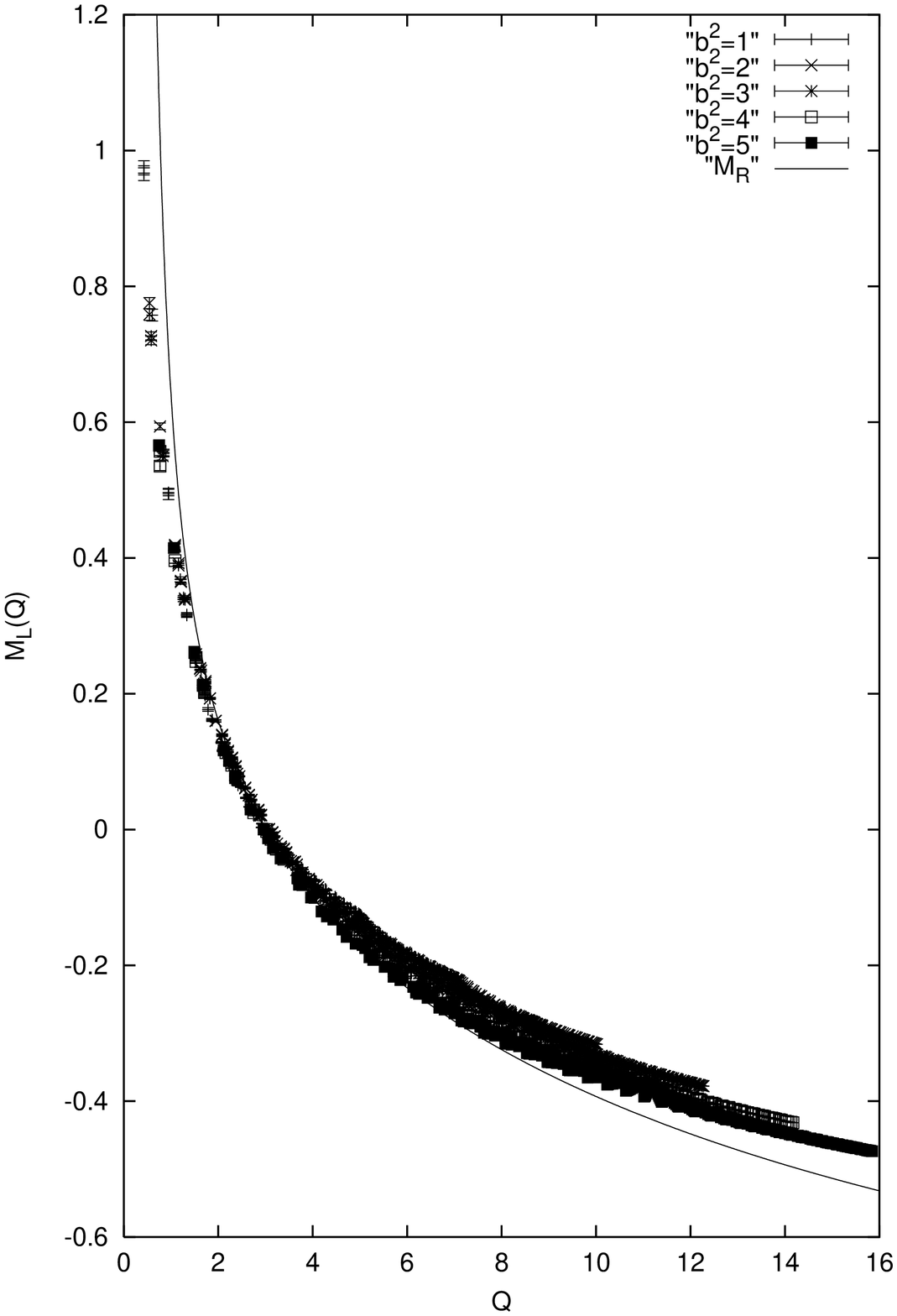}}
\caption{
Plot of the meson propagator for massless quarks  
as a function of
$b^2$. The $b^2=1$ data is at $N_c=37$, the $b^2=2,4$ data are at $N_c=41$
and the $b^2=3,5$ data are at $N_c=47$.
}
\label{propg0}
\end{figure}

Finally we look at massless quarks. To compare the lattice data to theory,
we need to pick a subtraction point. 
We used $\sqrt{Q^2}=3$. For any finite $\nc$, 
the scalar and pseudoscalar propagators are
identical for massless quarks. As explained earlier,
the common value of these reduced model propagators
ought to be compared with the continuum $M_R$ defined by:
\be
M_R(Q^2)= {1\over 2\pi} \sum_{n\ge 0}^\infty \left [ {r_n^2\over Q^2 +\mu_n^2} 
- {r_n^2\over 3^2 +\mu_n^2} \right ] .
\ee
The comparison for the subtracted meson propagator is shown 
in Fig. \ref{propg0} and is quite good.
The deviations that now appear at small $Q$ can be mostly attributed 
to finite $\nc$ effects which are sizable now because 
of the massless pseudoscalar mode.

\section{Summary and Conclusions}

The main objective of this paper was to introduce a 
new program whose aim is the numerical solution of 
planar QCD. Our first test was an application to two-dimensional
planar QCD. Much more can be done on this test case;
our results are just the very first steps. We hope
to carry out similar preparatory studies in four dimensions.

Our numerical experiment in two dimensions shows that
reduction can reproduce some correct numbers, but that
precise agreement would require a more substantial effort.
It should be kept in mind that QCD$_2$ in the planar limit 
is quite
challenging numerically. This model was chosen as our test case only 
because we have exact theoretical information about it. 
In four dimensions, 
the planar limit
of QCD might be approached more rapidly.

Perhaps surprisingly, two dimensions might be even 
harder than four dimensions in the reduced model framework.  
If all minima of the pure gauge action are correctly sampled,
$\nc\sim V^{\frac{1}{d}}$. Therefore, roughly, the number
of integration variables, $\nc^2$ decreases with the dimension
at fixed equivalent number of lattice sites. One might not
need more lattice sites in four dimensions than in two 
because infrared effects are stronger in two dimensions. 

The ultraviolet cutoff is increased by decreasing the 't Hooft scaled
coupling $e^2$, which is the same as taking our $b^2$ parameter to 
infinity. Numerically we have learned
that one needs larger $\nc$ values to see convergence to the planar
limit for larger values of $b^2$. The associated cost could be reduced by
improving the lattice action and our fermionic operators. 
Already from \cite{teper}, we learn that
simple improvement methods have a beneficial effect. After reduction, 
the relative benefit from improvement might be less than in regular
simulations. Of course, one needs to rethink what kind of algorithms are
best suited for reduced models. 

We have determined numerically that the eigenvalue distribution of the
parallel transporters around plaquettes has support in only a small
neighborhood of unity when we are at reasonably weak gauge coupling.
As a result, the spectrum of the Wilson Dirac operator develops a gap
around zero. This in turn would speed up the numerical algorithms
used to implement the lattice overlap propagators.

Reduction would also provide a means for evaluating $1/\nc$ corrections.
Thus, one might not only get numbers in the planar limit but also an
estimate for their accuracy in the context of real QCD. By going to the
Veneziano limit instead of to the 't Hooft limit, 
we might get meson widths in addition to the meson masses.
Recall that the momentum enters very differently
here so that one might even contemplate a direct analytical continuation in 
momentum
to physical values. The Veneziano limit, as a function of $N_f/\nc$,
would provide quantitative assessments of the valence approximation---something
of great value at this point in time. 

Furthermore, the old problem of dealing with complex actions numerically
might be more tractable in reduced models because the number of integration
variables seems to be so much smaller. This would open new ways to look at
$\theta$-dependence and at finite baryonic density. 

We conclude that pursuing our proposal further is a risk that is worth
taking and look forward to further developments in this direction. 

\begin{acknowledgments}
The research of H. N. was supported in part by DOE grant DE-FG02-01ER41165,
by a Guggenheim fellowship, and by a Lady Davis fellowship. 
H.~N.~would like to thank V.~Kazakov and the
entire theory group at Ecole Normale, M. Moshe and the entire theory group
at the Technion, and S. Nussinov and the entire theory group at Tel Aviv 
University for hospitality and support. H.~N. would also like to thank 
F. Berruto, L. Giusti, A. Gonz\'alez-Arroyo, 
V. Kazakov, C. Rebbi, and M. Teper for discussions related to the topics 
of this paper. 
We also wish to acknowledge W. Krauth and M. Staudacher for help
regarding numerical ways to obtain 't Hooft's exact solution. 
R.~N.~would like to
acknowledge the DOE grant DE-FG02-01ER41165 for computational support.
Part of the numerical work was done using the Linux cluster located
in the ITD of BNL. J.~K.~gratefully acknowledges access to the linux PC cluster
of W.~Pickett's condensed matter theory group at UC~Davis where some of the 
calculations were done.
\end{acknowledgments}

\newpage

\appendix

\section{}

In this appendix we describe the combination of numerical and analytical methods
used to produce the theoretical predictions for the meson correlators. Without
exception we are only dealing with the case of quarks of degenerate mass. Our
presentation is fairly detailed. 

't Hooft's Hamiltonian, $H$, is defined by:
\be
\label{thoofth}
(H\phi) (x) = \gamma (\frac{1}{x}+\frac{1}{1-x} ) \phi(x)- \int dy \frac {P}{(y-x)^2} 
[\phi(y)-\phi(x)] 
\ee

\subsection{Asymptotics of wave functions}

There is an $x\to 1-x$ symmetry under which the eigenfunctions
of $H$ will be either even or odd. We wish to determine the correct boundary
conditions on the wave functions and thus define the Hilbert space on which $H$
acts. We shall focus on the $x=0$ endpoint. The above reflection about $x=\frac{1}{2}$
determines then the boundary conditions at the other endpoint.

We write $\frac{P}{(y-x)^2} =-\frac{d}{dy}\frac{P}{y-x}$ 
in equation (\ref{thoofth}) and
perform an integration by parts. Then, we add and subtract $\phi^\prime (x)$ in
the integral, ending up with:
\begin{eqnarray}
(H\phi)(x)=\gamma\left ( \frac{1}{x}+\frac{1}{1-x} \right ) \phi(x) +
\frac{\phi(1)-\phi(x)}{1-x}-\frac{\phi(x)-\phi(0)}{x}\nonumber\\
-\frac{1}{2} \phi^\prime (x) \log\left (
\frac{1-x}{x} \right )^2 
-\int_0^1 dy \frac{\phi^\prime (y) -\phi^\prime (x)}{y-x}
\end{eqnarray}
We now assume asymptotics of the type
\begin{equation}
\phi (x) \sim x^\beta~~{\rm for~}x\to 0
\end{equation}
We plug this in and evaluate for $x\sim 0$. We use the asymptotic form
over the entire integration range inside the integral, but shall later make sure that
only the region of $y$ close to zero contributes to the final expression. 
Also assume that $0<\beta<1$. Then the two most singular terms at $x=0$
are contained in:
\begin{eqnarray}
\label{eqx1}
(H\phi)(x) \sim (\gamma -1) x^{\beta -1} +\frac{\beta}{2} x^{\beta -1} \log x^2
-\nonumber\\
-\beta x^{\beta -1} \int_0^{\frac{1}{x}} 
du \left [ \frac {u^{\beta-1}-1}{u-1} +\frac{1}{u+1}\right ]
+\beta x^{\beta -1} \log \frac{1}{x}
\end{eqnarray}
Integration variables were changed in the integral: $y=ux$. The term $\frac{1}{u+1}$
was added and subtracted. The subtracted integral is trivial and gives the second
term that contains a logarithm in eq. (\ref{eqx1}). The integral 
converges as the upper limit goes
to infinity. Thus, even if we stop the integration 
at a large finite number $A$ the answer would
be some number which will depend on $A$ only slightly. But, if $A$ is finite,
the range of $y$'s is from zero to $Ax$, so it is all in the asymptotic regime.
This is why we were justified in using the asymptotic expression for $\phi$ under the
integral sign, for what seemed to be the entire range. 
The integral (for $x=0$)
is done using the $\mu\to 1$ limit of
\bea
&\int_0^\infty du \frac{u^{\beta -1}- u^{\mu -1}}{u-1} +\int_0^\infty 
du\frac{u^{\mu -1}}{u+1}=\nonumber\\ 
&\pi (\cot \mu \pi - \cot \beta\pi ) +\frac{\pi}{\sin(\mu\pi)}
\eea
which is $-\pi \cot \pi\beta$. 

The terms containing logarithms cancel and to cancel the subleading singularities 
(which are still too strong to be matched by a term linear in $\phi$ - which
is what would be necessary for an eigenstate) we need:
\begin{equation}
\label{trans}
\gamma -1 +\pi\beta\cot\pi\beta=0
\end{equation} 
This formula can be found in 't Hooft's original paper \cite{thooft2d}. 

For light quarks, $\gamma\to 0$, and we get:
\begin{equation}
\beta \sim \frac{1}{\pi} \sqrt {3\gamma}~~{\rm for}~\gamma\to 0
\end{equation}

For massless quarks the asymptotic analysis needs to be redone.
In this case the Hamiltonian has the constant function as the lowest 
eigenstate. Thus, the action of $H$ on a function is determined by the
derivative of the latter.
\begin{eqnarray}
(H\phi)(x)=
\frac{\phi(1)-\phi(x)}{1-x}-\frac{\phi(x)-\phi(0)}{x}
-\frac{1}{2} \phi^\prime (x) \log\left (
\frac{1-x}{x} \right )^2 \nonumber\\
-\int_0^1 dy \frac{\phi^\prime (y) -\phi^\prime (x)}{y-x}
\end{eqnarray}
We assume
\begin{equation}
\phi^\prime(x) \sim x^\alpha ~~{\rm for}~x\to 0
\end{equation}
with $0<\alpha <1$ and check for consistency. This will give a condition on $\alpha$,
which will have a solution in the required interval. As before, we take the asymptotic
form throughout the integration range, and leave it to the end to check that the
contributions kept only came from the asymptotic regime. We find:
\begin{equation}
(H\phi)(x)= -\frac{1}{1+\alpha} x^\alpha +x^\alpha \log x -x^\alpha\int_0^{\frac{1}{x}}
du \frac{u^\alpha -1 }{u-1} +\cdots
\end{equation}
The $u$ integral is rewritten by adding and subtracting terms chosen for their 
large $u$ behavior, with the objective of isolating the contribution from the region
near the endpoint $y=0$. 
\begin{equation}
\int_0^{\frac{1}{x}}du \frac{u^\alpha -1 }{u-1}=
\int_0^{\frac{1}{x}}du \left [ \frac{u^\alpha -1 }{u-1} -u^{\alpha-1} +\frac{1}{u+1}\right ]
+\frac{1}{\alpha x^\alpha}+\log\frac{x}{1+x}
\end{equation}
Collecting terms in the integral and using
\begin{equation}
\int_0^\infty du u^{p-1}\frac{P}{1-u^q} =\frac{\pi}{q} \cot \frac {p\pi}{q}~~{\rm for}~
p<q
\end{equation}
we get
\begin{equation}
\int_0^\infty \frac{u^\alpha + u^{\alpha-1}-2}{1-u^2}=\pi \cot (\pi\alpha)
\end{equation}
As a result, we get
\begin{equation}
\\(H\phi ) (x)= -\frac{1}{\alpha}-\frac{x^\alpha}{\alpha+1} +\pi x^\alpha \cot \pi\alpha
+\cdots
\end{equation}
$\alpha$ is the unique solution between $0$ and $1$ of the equation:
\begin{equation}
\pi\cot\pi\alpha = \frac{1}{\alpha +1}
\end{equation}
This equation is the same as equation (\ref{trans}), extended to $1<\beta<2$, with 
$\beta=\alpha+1$ and with $\gamma=0$. Since the leading term is a constant,
we see that the next period of the cotangent in (\ref{trans}) provides the subleading
behavior. 

In summary we conjecture that the eigenfunctions for massive quarks
have the following expansion near $x=0$:
\begin{eqnarray}
\label{conj}
\phi_n^\gamma (x) \sim ( a^1_0 x^{\beta_1} + a^1_1 x^{\beta_1 +1} +\cdots)
+(a^2_0 x^{\beta_2} +a^2_1 x^{\beta_2 +1} +\cdots)\nonumber\\
+\cdots+
(a^k_0 x^{\beta_k} +a^k_1 x^{\beta_k +1} +\cdots)+\cdots
\end{eqnarray}
where $\beta_k$, $k=1,2,\cdots$ are, in increasing order, all the positive
solutions of equation (\ref{trans}). 
When the quark mass is taken to zero
the entire series associated with the first root is replaced by a constant. 

In the massless case we also can say that, for any excited eigenstate of $H$, 
$(H\phi_E)(x)=E\phi_E (x), E > 0$ we have,
\begin{equation}
\phi_E (x) = {\cal N} \left (1-\frac{\alpha E}{\alpha+1} 
x^{\alpha +1}\right ) +{\rm~terms~that~vanish~
faster~as~}x\to 0
\end{equation}
The value of $0< \alpha <1 $ can be easily calculated numerically. 
Note that the subleading correction at the endpoint holds in an ever decreasing
range as the energy of the state, $E$, increases.

In conclusion, the Hilbert space appropriate to the Hamiltonian for the case
of massless quarks consists of square integrable functions on the segment
obeying von Neumann boundary conditions $\phi^\prime (0)=\phi^\prime (1) =0$.
In the massive case, the boundary conditions were Dirichlet, $\phi(0)=\phi(1)=0$  
instead. 
Note that the internal Hilbert spaces corresponding to massless and massive quarks
are distinct, as one might have guessed would be appropriate for a manifestation
of spontaneous symmetry breaking in the planar limit.

\subsection{Massive quarks: $\gamma=1$}

Equation (\ref{trans}) simplifies when $\gamma=1$: its roots are $\beta_k = k-\frac{1}{2}$
with $k=1,2,\dots$. The double series in (\ref{conj}) collapses into a single series
and $\phi_n^1 (x)$ has the simple structure of $\sqrt{x}$ times a series in $x$
for the $x\to 0$ asymptotic regime. For the eigenvectors, there is a $x\to 1-x$
symmetry, so the structure should be
\begin{equation}
\phi_n^1 (x) = 2\sqrt{x(1-x)} f_n (x)
\end{equation}
where $f_n$ is either even or odd about the center of the segment $(0,1)$. 
It is convenient to map this segment to $(-1,1)$. 

For $\gamma=1$ the Hamiltonian is particularly simple:
\begin{equation}
(H\phi)(x)=-\int_0^1 dy \frac{P}{y-x} \phi^\prime (y)
\end{equation}
Under the change of variables $x\to 2x-1$, and still using a prime to denote
derivatives with respect to the new variable we get:
\begin{equation}
(H\phi )(x) =-2 \int_{-1}^1 dy \frac{P}{y-x}\phi^\prime (y)
\end{equation}
$\phi$ vanishes at the endpoints, so $\phi$=constant is not an eigenstate.
The eigenfunctions are defined over the segment $(-1,1)$
and have the structure (dropping the superscript denoting the
value of $\gamma$) 
\begin{equation}
\phi_n (y) = \sqrt{1-y^2} f_n (y)
\end{equation}

We shall use Chebyshev polynomials of the first ($T$) and second ($U$) kind to
parametrize $f_n$ \cite{peccei}. Some of their properties are listed below:
\begin{eqnarray}
\label{cheb}
&U_n (\cos \theta ) =\frac{\sin (n+1)\theta}{\sin\theta}\nonumber\\
&T_n (\cos\theta ) =\cos n\theta\nonumber\\
&\int_{-1}^1 dy \frac{P}{y-x} \frac{T_n (y)}{\sqrt{1-y^2}} = \pi U_{n-1} (x)
\nonumber\\
&\int_{-1}^1 dy \frac{P}{y-x} \sqrt{1-y^2} U_n (y) = -\pi T_{n+1} (x)
\nonumber\\
&\int_{-1}^1 dx \sqrt{1-x^2} U_n (x) U_m (x) =\frac{\pi}{2}\delta_{n,m}
\nonumber\\ 
&2(x^2-1)U_{m-1} (x) U_{n-1} (x) = T_{n+m} (x) - T_{|n-m|} (x)
\end{eqnarray}
The indices $n$ and $m$ above are non-negative and vary in the ranges in which the
equations make sense. 

We represent a given function $\phi (y)$ as
\begin{equation}
\phi (y) =\sqrt{1-y^2 } \sum_{n=0}^\infty a_n U_n (y)
\end{equation}
A short calculation then gives
\begin{equation}
\phi^\prime (y) =-\frac{1}{\sqrt{1-y^2}} \sum_{n=0}^\infty (n+1) a_n T_{n+1} (y)
\end{equation}
Using the properties of the Chebyshev polynomials listed in (\ref{cheb}) we find
\begin{eqnarray}
(H\phi ) (x) = 2\int_{-1}^1 dy \frac{P}{y-x} \frac{1}{\sqrt{1-y^2}} 
\sum_{n=0}^\infty
(n+1) a_n T_{n+1} (y) \nonumber\\
= 2\pi\sum_{n=0}^\infty (n+1) a_n U_n (x)
\end{eqnarray}
Suppose $\phi (x)$ is an eigenstate of $H$ with eigenvalue $E$, $(H\phi)(x) = E\phi(x)$.
Then,
\begin{equation}
E\sqrt{1-y^2} \sum_{n=0}^\infty a_n U_n (y) = 2\pi \sum_{n=0}^\infty (n+1) a_n U_n (y)
\end{equation}
One can proceed now by using orthogonality on one of the sides of the equation.
We choose to 
multiply the equation by $\sqrt{1-y^2} U_m (y)$ and integrate over $y$:
\begin{equation}
E\sum_{n=0}^\infty a_n \int_{-1}^1 dy (1-y^2) U_n (y) U_m (y) = \pi^2 (m+1) a_m
\end{equation}

We need to evaluate the integral on the left hand side. 
Using an identity in (\ref{cheb})
we find
\begin{equation}
\int_{-1}^1 (1-y^2 ) U_n (y ) U_m (y) =\frac{1}{2} \int_{-1}^1 [ T_{|n-m|} (y)
-T_{n+m+2} (y) ] \equiv \xi_{nm}=\xi_{mn}
\end{equation}
By $y\to -y$ we see that $\xi_{nm}=0$ for $n-m$ odd. Also,
\begin{equation}
\sum_{n=0}^\infty \xi_{mn} a_n =\frac{\pi^2}{E} (m+1) a_m
\end{equation}
We defined new coefficients 
\begin{equation}
b_n = \sqrt {n+1} ~a_n
\end{equation}
We also defined a new matrix $X$
\begin{equation}
\label{Xmat}
X_{mn} =\frac{1}{\sqrt{m+1}} \xi_{mn} \frac{1}{\sqrt{n+1}}
\end{equation}
We end up needing the eigenvalues of the symmetric matrix $X$:
\begin{equation}
\sum_{n=0}^\infty X_{mn} b_n = \frac{\pi^2}{E} b_m
\end{equation}
To calculate the entries $\xi_{mn}$ we go to angular variables: $y=\cos\theta$:
\begin{equation}
\xi_{mn} =\frac{1}{2} \int_0^\pi d\theta \sin\theta [\cos (n-m)\theta - \cos (n+m+2)\theta ]
\end{equation}
We need one elementary integral, for even integer $k$:
\begin{equation}
\int_0^\pi d\theta \sin\theta \cos k\theta = -\frac{2}{k^2 -1}
\end{equation}
For odd $k$ the integral vanishes.
This leads, for $n-m$ even, to
\begin{equation}
\xi_{mn} =\frac{1}{(n+m+2)^2 -1} -\frac {1} {(n-m)^2 -1}
\end{equation}
In particular, on the diagonal we have
\begin{equation}
\xi_{nn} = 1+\frac{1}{4(n+1)^2 -1}
\end{equation}
For $n-m$ odd the matrix element $\xi_{mn}$ vanishes. 
The matrix $X$ is given by:
\begin{equation}
X_{mn} = \frac{1}{\sqrt{(m+1)(n+1)}} \left [ \frac{1}{(n+m+2)^2 -1} - 
\frac {1}{(n-m)^2 -1} \right ]
\end{equation}
for $n-m$ even and $X_{mn}=0$ for $n-m$ odd. If $\lambda$ is an eigenvalue
of $X$, $E=\frac{\pi^2}{\lambda}$ is an eigenvalue of the 't Hooft Hamiltonian. 

For high states one can use the diagonal terms as an approximation. One gets
the asymptotic estimate
\begin{equation}
E_n \sim \pi^2 (n+1)
\end{equation}
with the state label $n$ starting at $n=0$. 

We need to determine the normalization convention on the infinite
vectors $(b_0, b_1,\dots,)$ that would make the eigenstates 
$\phi_E(x)$ ($H\phi_E = E\phi_E$) normalized
to unity by $\int_0^1 \phi_E^2 (x) =1$. 
$\phi_E (x)$ is parametrized by
\begin{equation}
\phi_E (x) =2\sqrt{x(1-x)}\sum_{n=0}^\infty \frac{b_n^E}{\sqrt{n+1}} U_n (2x-1)
\end{equation}
where $x$ is in the original range $(0,1)$. Introducing the expansion we find
\begin{equation}
1=\frac{\pi^2}{2E} \sum_{n=0}^\infty (b_n^E )^2
\end{equation}
where,
\begin{equation}
\sum_{n=0}^\infty X_{mn} b_n =\frac{\pi^2}{E} b_m
\end{equation}

We are now ready to compute the residues for $\gamma=1$. We need to calculate 
\begin{equation}
\rho_n^1 =\int_0^1 \frac{dx}{x} \phi_n (x)
\end{equation}
Here, $\gamma$ was set to unity and the eigenfunction is assumed correctly
normalized. Changing variables, we arrive at
\begin{equation}
\rho_n^1 = \sum_{k=0}^\infty a_k^{(n)} I_k
\end{equation}
with
\begin{equation}
I_k = \int_0^\pi d\theta \frac{\sin\frac{\theta}{2}}{\cos\frac{\theta}{2}}
\sin{(k+1)\theta}
\end{equation}
We change the integration variable $\theta$ to $\pi-\theta$ 
and average the expressions.
Then, for $k$ even we have to calculate
\begin{equation}
I_k = \int_0^\pi d\theta\frac{\sin(k+1)\theta}{\sin\theta}
\end{equation}
Use
\begin{equation}
\frac{\sin(k+1)\theta}{\sin\theta}=\sum_{j=0}^k e^{\imath (k-2j)\theta}
\end{equation}
to conclude that for even $k$ only the $j=\frac{k}{2}$ term in the sum
makes a contribution, giving
\begin{equation}
I_k =\pi
\end{equation}
For $k$ odd 
\begin{equation}
I_k = -\int_0^\pi d\theta\frac{\sin(k+1)\theta}{\sin\theta}\cos\theta
\end{equation}
By the same technique as above we get non zero contributions only from 
$j=\frac{k\pm 1}{2}$ (for odd $k$, $k\ge 1$). Hence, for $k$ odd:
\begin{equation}
I_k =-\pi
\end{equation}
Only the absolute value of $\rho_n^1$ is determined, since the sign of the
wave function is not fixed by normalization.
The final answer is:
\begin{equation}
\label{resga}
|\rho_n^1| =\pi \Big | \sum_{k\ge 0,~k-n=~{\rm even}} 
\frac{b_k^{(n)}}{\sqrt{k+1}}\Big |
\end{equation}
Here, the normalization condition is:
\begin{equation}
\label{normga}
\sum_{k\ge 0,~k-n=~{\rm even}} ( b_k^{(n)})^2 =\frac{2 E^{(n)}}{\pi^2}
\end{equation}

In each sector the eigenvalues were
obtained numerically by
diagonalizing a truncation of the infinite matrix $X$ of (\ref{Xmat}).
The residues were
obtained using (\ref{resga}) with the eigenfunctions of the truncated
matrix $X$ normalized
according to (\ref{normga}). In each sector,  
a matrix of size $2000\times 2000$ was diagonalized
and this produced accurate estimates of the lowest $1000$ eigenvalues
in that sector.
The lowest pole contributing to the pseudoscalar is at $\mu_0=2.69713$
and the lowest pole contributing to the scalar is at $\mu_1=4.16036$.
Asymptotically, the eigenvalues are given by $\mu_n^2=\pi^2 n + {3\over 4}
\pi^2 + {\cal O}\frac{1}{n}$.
The residues, $r_n^2$, reach an asymptotic value of $\pi^2$. The even residues
approach this number monotonically from above ($r_0^2=11.7596864$)
and the odd residues approach this number monotonically from below 
($r_1^2=9.03541704$).

\subsection{Massless quarks}

We now turn to the evaluation of eigenvalues and residues in the case of
massless quarks. The correct boundary conditions for the massless case are 
von Neumann so we choose to diagonalize the Hamiltonian in the basis of 
cosine functions.
\begin{equation}
\langle x | n \rangle = \sqrt{2} \cos {\pi n x} ~~~n\ge 1;~~~~\langle x | 0 \rangle =1
\end{equation}
The states are numbered, starting with 0. The ground state is known exactly and by
keeping $n\ge 1$ we are in a space orthogonal to it. Just as before, even and odd $n$
states do not mix, so we can diagonalize the Hamiltonian in each of these
subspaces separately. 
\begin{eqnarray}
\label{Hmat}
\langle n | H | m \rangle =& \int_0^1 dx \int_0^1 dy \int_0^\infty t dt e^{-t|x-y|}
[\cos (\pi n x ) - \cos ( \pi n y ) ] \nonumber \\
&[\cos (\pi m x ) - \cos ( \pi m y ) ]
\end{eqnarray}
To do the integrals we introduce, 
with no restriction on the integers $n,m$ the integral
\begin{equation}
I_{n,m} = \frac{1}{2}\int_0^1 dx \int_0^1 dy \int_0^\infty t dt e^{-t|x-y|}
[e^{\imath \pi n x} - e^{\imath \pi n y}]
[e^{\imath \pi m x} - e^{\imath \pi m y}]
\end{equation}
By changing variables $x\to 1-x$ and $y\to 1-y$ we see that
\begin{equation}
I_{n,m} = (-1)^{n+m} I^*_{n,m}
\end{equation}
We now restrict ourselves to even $n+m$. 
This is enough to get all nonzero matrix elements
$ \langle n | H | m \rangle$. In this case we learn that $I_{n,m}$ is real. 
\bea
&I_{n,m}+I_{n,-m} =\nonumber\\
&\int_0^1\int_0^1 \int_0^\infty t dt e^{-t|x-y|}
[e^{\imath \pi n x} - e^{\imath \pi n y}]
[\cos ( \pi m x ) - \cos (\pi m y)]
\eea
But, since $I_{n,m}$ is real we can take the real part to prove
\begin{equation}
\langle n | H | m \rangle=I_{n,m}+I_{n,-m}
\end{equation}
We need only $I_{n,m}$ for $n+m$ even. Using the symmetry under exchange
of $x$ and $y$ we get
\begin{equation}
I_{n,m} = \int_0^1 dx \int_0^x dy  \int_0^\infty t dt e^{-t(x-y)}
[e^{\imath \pi n x} - e^{\imath \pi n y}]
[e^{\imath \pi m x} - e^{\imath \pi m y}]
\end{equation}

Using the above, and restricting in the end to $n,m \ge 1$ and 
even $n+m$ we get:
\begin{eqnarray}
&\langle n | H | m \rangle = \pi^2 n \delta_{n,m} + 2\int_0^\infty t dt\nonumber\\
& \left \{ [1+(-1)^{n-1}e^{-t}] \left [
\frac{1}{t^2+\pi^2 n^2}+\frac{1}{t^2+\pi^2 m^2}-
\frac{\pi^2(n^2 + m^2 )}{(t^2+\pi^2 n^2)(t^2+\pi^2 m^2)} \right ] \right . \nonumber\\
&-\left . (1-e^{-t} ) \left [ \frac{1}{t^2+\pi^2 (n+m)^2} +\frac{1}{t^2 +\pi^2 (n-m)^2}\right ]\right \}
\end{eqnarray}

The matrix elements of $H$ can be expressed in terms of sine-integral and cosine-integral
functions:
\begin{eqnarray}
\label{eqx2}
&{\rm For~}n\ne m,~{\rm with}~n+m~{\rm even}:\nonumber\\
&\langle n | H | m \rangle = \log \left ( \frac{n^2 -m^2 }{nm}\right )^2 
+\frac{n^2+m^2}{n^2-m^2} \log \left ( \frac{m}{n}\right )^2 \\
&+\frac{4}{n^2 - m^2} [ n^2 ci(\pi n) - m^2 ci (\pi m)]
-2\{ ci[\pi(n+m)] + ci [\pi(n-m)]\}\nonumber
\end{eqnarray}
\begin{eqnarray}
\label{eqx3}
&{\rm For~}n= m:\nonumber\\
&\langle n | H | n \rangle = \pi^2 n -2\log n +\log\frac{4}{\pi^2} -2(1+\gamma )\nonumber\\
&+2[\pi n si (\pi n)+(-1)^n] 
+ 4 ci(\pi n) - 2 ci(2\pi n)
\end{eqnarray}
One needs to distinguish between Si, Ci and si, ci. 
Here are the definition for the functions
used in equations (\ref{eqx2}) and (\ref{eqx3}): 
\begin{eqnarray}
&si(x)=-\int_x^\infty dt \frac{\sin t}{t}=-\frac{\pi}{2}+\int_0^x dt \frac{\sin t}{t}=
-\frac{\pi}{2}+Si(x)\\
&ci(x)=-\int_x^\infty dt\frac{\cos t}{t}=Ci (x) 
\end{eqnarray}
The asymptotics go as follows:
\begin{eqnarray}
&{\rm For~integer~and~large~positive~}k:\nonumber\\
&ci(\pi k) = \frac{(-1)^{k-1}}{(\pi k)^2} +{\cal O} \left ( \frac{1}{(\pi k )^4}\right )\nonumber\\
&si(\pi k) =\frac{(-1)^{k-1}}{\pi k} +{\cal O} \left ( \frac{1}{(\pi k )^3}\right )
\end{eqnarray}

To find the residues in the case of massless quarks one needs to take the
$\gamma\to 0$ limit on the massive formula:
\begin{equation}
\rho_n^\gamma=\sqrt{\gamma} \int_0^1 dx \frac{\phi_n^\gamma (x)}{x}
\end{equation}
$\phi_n^\gamma (x)$ is the $n$-th (ordered by eigenvalue and starting from
$n=0$) normalized eigenstate of the Hamiltonian with mass parameter $\gamma$. 

The entire contribution comes from the lower end of the integral, because only
from there does one get a singularity in $\gamma$ that can compensate the vanishing
prefactor. Thus, only the asymptotic behavior of the wave function is needed.
It is given by:
\begin{equation}
\phi_n^\gamma (x) = A_n^\gamma x^{\beta(\gamma)} + 
{\rm~terms~that~vanish~faster~as~}x\to 0
\end{equation}
Hence,
\begin{equation}
\label{resgb}
\rho_n^0 =\frac{\pi}{\sqrt{3}} A_n^0
\end{equation}
In general, $A_n^0$ is difficult to obtain, as it is fixed by the normalization
of the wave function and depends on its values throughout the interval. But,
for $n=0$ we know that, at small $\gamma$, $\phi_0 (x)$ tends to $1$. Hence,
$A_0^0=1$ and we get
\begin{equation}
\rho_0^0=\frac{\pi}{\sqrt{3}}
\end{equation}
For arbitrary $n$ we need to compute numerically the $n$-th wave function
for the massless case, normalize it, and get $A_n^0$ from its values at the
end point.
\begin{equation}
A_n^0=\phi_n^0(0)
\end{equation}
A sign ambiguity remains, as the sign of the wave function remains undetermined
by the normalization condition. But, only $(\rho_n^0)^2$ enters in the amplitude
so everything is well determined.

The eigenvalues 
were obtained from a truncation of $H$ in (\ref{Hmat}) to  
size $1000\times 1000$ in each sector. The residues are obtained
from the end point of the normalized eigenvector as given by (\ref{resgb}).
The lowest non-zero eigenvalue is a scalar meson (having an 
antisymmetric wave function) at 
$\mu_1=2.4233$ and its residue is $r_1^2=7.58941916$. Asymptotically,
the eigenvalues
are described by $\mu_n^2=\pi^2 n - 2\log(n) + O(1)$. The constant term
is close to ${1\over 2}\pi^2$ but due to the $\log(n)$ term 
it is hard to determine it accurately. 
The residues are relatively difficult to estimate due to 
slow convergence. They asymptote to $r_n^2=\pi^2$ and are monotonic in $n$.

\subsection{Light versus heavy quarks}

We wish to determine whether $\gamma=1$ should be viewed as a light quark
mass or as a heavy quark mass. To do this we first find the leading behavior
of the pion mass for very small quark masses. This behavior has the typical
structure induced by spontaneous chiral symmetry breaking. We then compare
the leading chiral approximation to the pion mass with the exact value.
If the quarks are light the approximation should work well numerically. 
If the quarks are heavy the approximation should be off. We also compute
the pion mass for very heavy quarks. Again we compare the approximate
expression to the exact value. 

To compute the pion mass ($n=0$ state) to leading order in the assumed small
quark mass we start from:
\begin{eqnarray}
(\mu_0^\gamma)^2 =
\int_0^1 dx \phi_0^\gamma (x) (H\phi_0^\gamma )(x) = \gamma\int_0^1 [\phi_0^\gamma (x)]^2
\left ( \frac{1}{x}+\frac{1}{1-x} \right )\nonumber\\
 -\int_0^1 dx\int_0^1 dy
\frac{P}{(y-x)^2} \phi_0^\gamma (x)[\phi_0^\gamma (y) -\phi_0^\gamma (x)]
\end{eqnarray}
Symmetrize both terms, the first under parity $x\to 1-x$, 
the second under interchange of $x$ with $y$ to get
\begin{equation}
(\mu_o^\gamma)^2=2\gamma\int_0^1 \frac{[\phi_0^\gamma (x)]^2}{x}+\frac{1}{2}
\int_0^1 dx\int_0^1 dy 
\left [ \frac{\phi_0^\gamma (y) -\phi_0^\gamma (x)}{y-x}\right ]^2
\end{equation}
As $\gamma\to 0$ the first term is dominated by
the endpoint contribution. Actually, also the second term is dominated
by the end point contribution. As before, we use the asymptotic behavior throughout
the integration range, but ascertain at the end that we were dominated by
the endpoints. From the first term we get a contribution:
\begin{equation}
\frac{\pi}{\sqrt{3}}\sqrt{\gamma}+{\cal O}(\gamma )
\end{equation}
The second term requires more work:
\begin{equation}
\frac{1}{2} \int_0^1 dx \int_0^1 dy \left [ \frac{y^\beta - x^\beta}{y-x} \right ]^2
=\frac{1}{2}\int_0^1 dx x^{\beta-1}\int_0^{\frac{1}{x}}du 
\left [ \frac{u^\beta -1}{u-1} \right ]^2 
\end{equation}
The $u$-integral is done extending the upper limit to infinity. The correction needed
to account for this approximation will be of higher order in $\beta$. 
The $u$ integral is done as follows:
Write $\left [ \frac{1}{u-1}\right ]^2 = -\frac{d}{du} \frac{1}{u-1}$ and do
the integral by parts to obtain
\begin{eqnarray}
\int_0^\infty du \left [ \frac{u^\beta -1}{u-1} \right ]^2 =
-1-2\beta\int_0^\infty
du \frac{u^{2\beta -1}-u^{\beta-1}}{1-u}=\nonumber\\
-1-2\beta\pi[\cot(2\beta\pi)-\cot(\beta\pi)]=
\frac{2}{3}(\beta\pi)^2 +{\cal O} (\beta^4 )
\end{eqnarray}
Doing the remaining $x$-integral we learn that the double integral gives a contribution
equal to that of the first term.

In summary we get, converting to physical units:
\begin{equation}
\label{mpi}
m_\pi^2 = 2 m_q |g| \sqrt{ \frac { \pi N}{3} } +{\cal O}(m_q^2)
\end{equation}
This is the result found by 't Hooft \cite{thoofterice}. Equation (\ref{mpi})
can be rewritten in variables rendered unitless by using $\frac{g^2 N}{\pi}$
as a mass-square scale as
\begin{eqnarray}
&\mu_\pi^2 = \sqrt {\gamma} \frac{2\pi}{\sqrt{3}}\nonumber\\
&\frac{m_\pi}{m_q}=\frac{\mu_\pi}{\sqrt{\gamma}}=\sqrt{\frac{2\pi}{\sqrt{3}}} \gamma^{-\frac{1}{4}}
\approx \frac{1.90}{\gamma^{\frac{1}{4}}}
\end{eqnarray}
For $\gamma=1$ we found numerically that $\frac{m_\pi}{m_q}=2.7$
establishing that $\gamma=1$ corresponds to quite light quarks, but still
the ${\cal O} (m_q^2 )$ corrections to eq. (\ref{mpi}) 
make a substantial contribution. 

The pion is, by definition, the lowest energy eigenstate of 't Hooft's
Hamiltonian in the sector of functions symmetric under $x\to 1-x$. 

For $\gamma$ large the first term in 't Hooft's Hamiltonian dominates,
so the pion wave function will try to minimize its contribution by being
concentrated around the point $x=\frac{1}{2}$. Making the wave function
too narrow incurs a price from the second term. This suggests a variational
estimate for the leading and subleading
order in $\frac{1}{\gamma}$. The trial wave function is
\begin{equation}
\phi (x) = \left ( \frac{\lambda}{\pi}\right )^{\frac{1}{4}}
\left [ e^{-\frac{\lambda}{2}(x-\frac{1}{2})^2}-e^{-\frac{\lambda}{8}}\right ]
\end{equation}
$\phi(x)$ is normalized to unity up to corrections that are exponentially
small in $\lambda$. 

The expectation value of 't Hooft's Hamiltonian in the state $\phi$ is:
\begin{equation}
\langle \phi |H|\phi\rangle =\gamma\int_0^1 \left ( \frac{1}{x}+\frac{1}{1-x} \right )
\phi^2 (x) 
+\frac{1}{2} \int_0^1 dx \int_0^1 dy \left [ \frac{\phi(y)-\phi(x)}{y-x} \right ]^2
\end{equation}

Contributions from the endpoints of the integral in the first term are exponentially
small in $\lambda$ and will be neglected, under the assumption (to be later
justified) that $\lambda$ diverges as $\gamma\to\infty$. Thus we can calculate the
contribution of the first term by infinite range Gaussian integration, obtaining
\begin{equation}
4\gamma + \frac{8\gamma}{\lambda} + {\cal O} (\lambda^{-2})
\end{equation}
The second term is
\begin{equation}
\frac{1}{2} \sqrt{\frac{\lambda}{\pi}} \int_0^1 dx \int_0^1 dy 
\left [ \frac{e^{-\frac{\lambda}{2} (x-\frac{1}{2} )^2 } -
e^{-\frac{\lambda}{2} (y-\frac{1}{2} )^2 }}{x-y} \right ]^2 
\end{equation}
The integral (excluding the prefactor) has a finite $\lambda\to\infty$ limit:
\begin{equation}
\int_{-\infty}^{\infty} du \int_{-\infty}^{\infty }dv 
\left ( \frac{e^{-\frac{u^2}{2}} - e^{-\frac{v^2}{2}}}{u-v} \right )^2
\end{equation}
The expression $(u-v)^2$ in the denominator is represented by an integral over $t$
of $e^{-t(u-v)^2}$ 
from zero to infinity. The $u,v$ integrals of each one of the terms can be
done by Gaussian integration, and calculating the appropriate $2\times 2$
determinants one finds that the above integral is given by:
\begin{equation}
2\pi\int_0^\infty \left [ \frac{1}{\sqrt{t}}- \frac{1}{\sqrt{t+\frac{1}{4}}} \right ] 
= 2\pi
\end{equation}

Hence
\begin{equation}
\langle H \rangle \sim 4\gamma +\frac{8\gamma}{\lambda} + \sqrt{\pi\lambda}
\end{equation}
Extremizing on $\lambda$ gives
\begin{equation}
\lambda=\left ( \frac{16}{\sqrt{\pi}} \right )^{\frac{2}{3}} \gamma^{\frac{2}{3}}
\end{equation}
Thus, indeed $\lambda$ is large for $\gamma$ large and the wave function is close
to a delta function around $x=\frac{1}{2}$. For the unitless pion mass square we
find: 
\begin{equation}
\mu_{\pi}^2 \sim 4\gamma +(16\pi )^{\frac{1}{3}}\frac{3}{2} \gamma^{\frac{1}{3}}
\end{equation}

For large $\gamma$ we expect therefore
\begin{equation}
\frac{m_\pi}{m_q} \sim 2 (1+0.69 \gamma^{-\frac{2}{3}} )
\end{equation}

Using this formula blindly for $\gamma=1$ gives $\frac{m_\pi}{m_q}\sim 3.4$
instead of the true value of $2.7$. We see that the true value is midway
between the leading answer corresponding to light pions (1.9) and the
leading plus subleading expression valid for heavy quarks. Thus, $\gamma=1$
corresponds to intermediate gauge coupling (in units of quark mass).

\end{document}